\documentclass[aip,jcp,amsmath,amssymb,preprint]{revtex4-1}
\pdfoutput=1

\usepackage{graphicx}
\usepackage{dcolumn}
\usepackage{bm}
\usepackage[utf8x]{inputenc}
\usepackage[T1]{fontenc}

\newcommand  {\hb}	 {\hat{\textbf{b}}} 
\newcommand  {\hp} {\hat{\textbf{p}}} 
\newcommand  {\hs} {\hat{\textbf{s}}}
\newcommand  {\vecr} {\textbf{r}}  
\newcommand  {\eij} {\epsilon_{ij}}
\newcommand  {\apa} {a_{\text{p}}} 
\newcommand  {\aap} {a_{\text{ap}}} 
\newcommand  {\bhh} {b_{\text{hh}}} 
\newcommand  {\bhp} {b_{\text{hp}}} 
\newcommand  {\bpp} {b_{\text{pp}}}

\newcommand  {\Emin}{E_{\text{min}}} 
\newcommand  {\Tf}      {T_{\text{f}}}

\newcommand  {\cf}      {c_{\text{f}}}
\newcommand  {\cs}      {c_{\text{s}}}

\newcommand  {\e}         {\text{e}}
\newcommand  {\Pacc}	{P_{\text{acc}}} 

\newcommand  {\thalf}      {t_{1/2}}
\newcommand  {\Cv}      {C_V}
\newcommand  {\Cvm}      {C_{V,\max}}

\newcommand   {\ev}[1]          {\langle#1\rangle}
\newcommand*{\citen}[1]{%
  \begingroup
    \romannumeral-`\x 
    \setcitestyle{numbers}%
    \cite{#1}%
  \endgroup   
}
\newcommand{\pDer}[2]{\frac{\partial #1}{\partial #2}}

\begin{document}

\title[]{Thermodynamics of amyloid formation and the role of intersheet interactions}

\author{Anders Irb\"ack}
\email{anders@thep.lu.se}
\affiliation{Department of Astronomy and Theoretical Physics, Lund University, S\"olvegatan 14A, SE-223 62 Lund, Sweden}

\author{Jonas Wess\'en}%
\email{jonas.wessen@thep.lu.se}
\affiliation{Department of Astronomy and Theoretical Physics, Lund University, S\"olvegatan 14A, SE-223 62 Lund, Sweden}

\date{\today}

\begin{abstract}
  The self-assembly of proteins into $\beta$-sheet-rich amyloid 
  fibrils has been observed to occur with sigmoidal kinetics, 
  indicating that the system initially is trapped 
  in a metastable state. Here, we use a minimal 
  lattice-based model to explore the thermodynamic forces 
  driving amyloid formation in a finite canonical ($NVT$) system.
  By means of generalized-ensemble Monte Carlo techniques
  and a semi-analytical method, the thermodynamic properties  
  of this model are investigated for different sets of intersheet 
  interaction parameters. When the interactions support
  lateral growth into multi-layered fibrillar structures, an
  evaporation/condensation transition is observed, between a 
  supersaturated solution state and a thermodynamically 
  distinct state where small and large fibril-like species 
  exist in equilibrium. Intermediate-size aggregates are 
  statistically suppressed. These properties do not hold if 
  aggregate growth is one-dimensional.
\end{abstract}


\maketitle

\section{\label{sec:intro}Introduction}

The formation of amyloid fibrils is currently an intensely studied 
phenomenon.\cite{Chiti:06,Knowles:11,Hard:14} 
Protein aggregates of this type are found in pathological deposits in
several human diseases, but also with functional roles. In addition,
they possess interesting mechanical properties, stemming 
from their characteristic ordered cross-$\beta$ organization.
Insight into the mechanisms of amyloid formation has been 
gained from kinetic profiles, as measured primarily
by using thioflavin T (ThT) fluorescence.\cite{Hellstrand:10} 
In particular, it has been shown that kinetic data for a broad 
range of systems can be 
well described in terms of a few basic mechanisms for the 
nucleation and growth of fibrils, through a rate-equation 
approach.\cite{Knowles:09} This approach can reveal some 
general properties of intermediate species participating in the
aggregation process, and has proven useful for related self-assembly 
phenomena as well.\cite{Ferrone:85,Flyvbjerg:96} 

Structure-based modeling of amyloid formation is a 
challenge to implement, due to the wide range of spatial and 
temporal scales involved. Hence, all-atom computer simulations 
with explicit solvent have focused on characterizing 
monomeric forms and  early aggregation events.\cite{Straub:10} 
By using coarse-grained models, at various levels of 
resolution, it has been possible to study the formation and stability of
larger assemblies~\cite{Auer:08,Junghans:08,Irback:08,Li:08,
LiMS:08,Bellesia:09,Lu:09,Wang:10,Auer:10,Kashchiev:10,Friedman:10,Rojas:10,
Urbanc:10,Kim:10,Cheon:11,Linse:11,Baiesi:11,Carmichael:12,Bieler:12,
Smaoui:13,Ni:13,Zheng:13,DiMichele:13,Abeln:14,MorrissAndrews:14,Saric:14,Assenza:14} 
and also get insight into the thermodynamic forces at play in amyloid
formation.\cite{Zhang:09,Auer:11,Schmit:11,Auer:14} However,   
to map out the thermodynamics of amyloid formation 
as a function of control 
parameters such as temperature and concentration is 
computationally demanding even in simple models. 

In this article, we use cluster~\cite{Swendsen:87} 
and generalized-ensemble~\cite{Berg:91,Hansmann:93,Wang:01a}  
Monte Carlo (MC) techniques, supplemented with a semi-analytical 
approximation, to investigate the thermodynamics of a 
minimal model for amyloid formation.\cite{Irback:13}  
We study this model for three different choices of 
intersheet interaction parameters. The first
choice leads to aggregates with at most two layers, and 
therefore an essentially 1D growth. The second choice
permits aggregates with more than two layers to form, but 
odd-layered aggregates are energetically suppressed. This choice,
inspired by evidence that the core of amyloid fibrils often 
has a pairwise $\beta$-sheet 
organization,\cite{Sawaya:07,Fitzpatrick:13} 
leads to a stepwise, quasi-2D growth. In the third and final 
case, odd-layered aggregates are not suppressed, 
which opens up for 2D growth, although slower laterally 
than longitudinally. Using $NVT$ ensembles ($N$ is the 
number of peptides, $V$ is volume, $T$ is temperature), 
we investigate the equilibrium properties of these three
systems as a function of  $T$ and the concentration $c=N/V$.
In addition, we study the relaxation of the systems 
in MC simulations under fibril-favoring conditions,  
starting from random initial states.

\section{\label{sec:methods}Methods}
\subsection{Model}

We use a minimal model for peptide 
aggregation where each 
peptide $i$ is represented by a unit-length stick 
centered at a site, $\vecr_i$, on a periodic cubic lattice 
with volume $V=L^3$.\cite{Irback:13} 
It is thus assumed that
the internal dynamics of the peptides are fast, compared
to the timescales of fibril formation, and can be   
averaged out. The systems studied consist of $N$ identical 
peptides or sticks. 
Two peptides cannot simultaneously occupy 
the same site. The orientation of a peptide is specified 
by two perpendicular lattice unit vectors, $\hb_i$ and 
$\hp_i$, yielding a total of 24 orientational states. The
$\hb_i$ vector represents the N-to-C backbone direction,
whereas $\pm\hp_i$ are the directions in which the peptide 
can form intrasheet interactions. The vectors  
$\pm\hs_i=\pm\hb_i\times\hp_i$ represent sidechain 
directions, in which the peptide can form intersheet interactions. 
Throughout the article, we assume units in which the lattice 
spacing, the peptide mass and Boltzmann's constant have the 
value one.

\begin{figure}[t]
\centering
  \includegraphics[width=8cm]{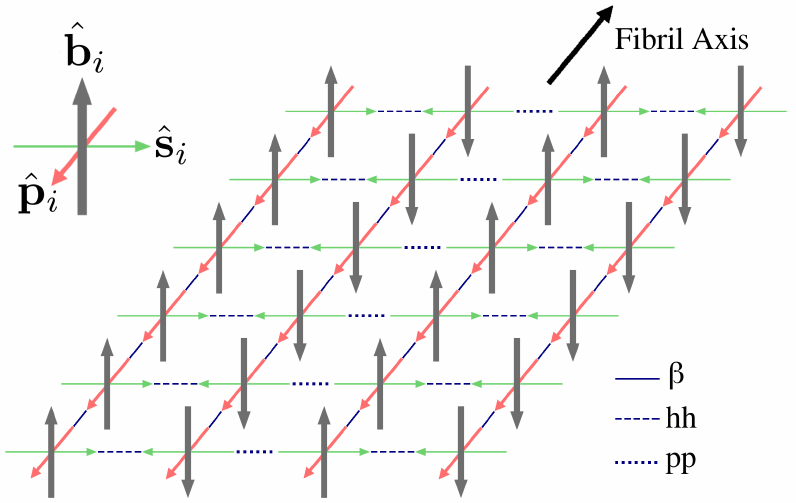}
  \caption{
  Example of a rectangular aggregate with length $l=6$ and width 
  $w=4$. Along the fibril axis, the peptides are bound together by parallel 
  $\beta$ interactions (solid blue, $\beta$). The four layers are connected 
  either via hh interactions (blue dashes, hh), or via pp interactions 
  (blue dots, pp). For potential B, pp interactions are weaker than 
  hh interactions, which favors even-layered aggregates. The symmetric potential   
  A assigns equal energy  to pp, hh and hp bonds. With potential C, hp and pp 
  interactions are missing, 
  which makes lateral growth beyond two layers impossible.
  \label{fig:new}}
\end{figure}

The interaction energy is taken to have a pairwise additive 
form, $E=\sum_{i<j}\eij$, where $\eij\le0$. The interaction
geometry is illustrated in Fig.~\ref{fig:new}. Consider an
arbitrary pair $i$ and $j$ of peptides and let  
$\vecr_{ij}=\vecr_j-\vecr_i$. The peptides 
interact ($\eij\ne0$) only if (i) they are nearest neighbors on the 
lattice ($|\vecr_{ij}|=1$), (ii) their backbone vectors are
aligned either parallel or antiparallel to each other 
($|\hb_i\cdot\hb_j|=1$), and 
(iii) $\hb_i\cdot\vecr_{ij}=\hb_j\cdot\vecr_{ij}=0$.
The interaction that takes  
place when these conditions are met can be of one 
of three types, depending on the relative orientation 
of the peptides:
\begin{enumerate} 
\item The interaction is of intrasheet type if both $\hp_i$ and $\hp_j$ 
equal $\pm\vecr_{ij}$, and $\eij$ is then  
given by
\begin{gather}     
\label{eq:intrasheet}
\eij=
\begin{cases}
  -(1+\apa) & \text{if}\ \hb_i\cdot\hb_j=\hp_i\cdot\hp_j=1 \\
  -(1+\aap) & \text{if}\ \hb_i\cdot\hb_j=\hp_i\cdot\hp_j=-1 \\
  -1 & \text{otherwise}
\end{cases}\\
\begin{split}
\nonumber
\qquad\qquad\qquad\qquad\qquad(|\vecr_{ij}|=|\hb_i\cdot\hb_j|=1,\ \hp_i, \hp_j=\pm \vecr_{ij}) 
\end{split}
\end{gather}    
where the first two cases represent parallel and
antiparallel $\beta$-sheet structure, respectively.
\item The interaction is of intersheet type if neither $\hp_i$ nor 
$\hp_j$ equals $\pm\vecr_{ij}$, which implies that both 
$\hs_i$ and $\hs_j$ equal $\pm \vecr_{ij}$. 
The $+\hs$ and $-\hs$ sides of a peptide, denoted by h and p, 
respectively, are assumed to have different interaction 
properties. The $+\hs$, or h, side is taken as  more sticky
or hydrophobic.  The pair potential is given  
by  
\begin{gather}     
\label{eq:intersheet}
\eij=
\begin{cases}
  -(1+\bhh) & \text{if}\ \hs_i=-\hs_j=\vecr_{ij} \\
  -(1+\bhp) & \text{if}\ \hs_i=\hs_j\\
  -(1+\bpp) & \text{if}\ -\hs_i=\hs_j=\vecr_{ij}
\end{cases}
\\
\begin{split}
\nonumber
\qquad\qquad\qquad\qquad\qquad(|\vecr_{ij}|=|\hb_i\cdot\hb_j|=1,\ \hs_i, \hs_j=\pm \vecr_{ij}) 
\end{split}
\end{gather}    
and is assumed lowest when the two h sides face each other ($\bhh\ge\bhp,\bpp$). 
\item If the interaction is of neither of these two types ($|\vecr_{ij}|=|\hb_i\cdot\hb_j|=1,
\hp_i\cdot\hp_j=\hs_i\cdot\hs_j=0$), the pair potential is set to $\eij=-1$.
\end{enumerate}

The intersheet interactions must be weak compared to 
the intrasheet interactions for elongated fibril-like 
aggregates to form, but are nevertheless important.    
To assess the role played by the intersheet interactions, 
we study the model using three potentials  
A, B and C, which differ in the choice of the 
parameters $\bhh$, $\bhp$ and $\bpp$ (Table~\ref{tab:b}). 
The intrasheet parameters 
$\apa$ and $\aap$ are the same in all 
three cases, namely $\apa=5$ and $\aap=3$.  

\begin{table}[b] 
  \caption{\label{tab:b} Our three choices of the intersheet interactions 
  parameters $\bhh$, $\bhp$ and $\bpp$ (Eq.~\ref{eq:intersheet}),
  and the corresponding growth behavior of aggregates.}
  \centering
  \begin{ruledtabular}
	\begin{tabular}{ccccl}
	Potential & $\bhh$ & $\bhp$ & $\bpp$ & Growth\\
	\hline
	A & 0.5 & 0.5 & 0.5 & 2D\\
	B & 1 & 0 & 0 & quasi-2D\\
	C & 1 & $-1$ & $-1$ & 1D\\
	\end{tabular}
  \end{ruledtabular}
\end{table}

Our previous study of this model was carried out using
potential B.\cite{Irback:13} With this potential, it was shown
that aggregates grow in a stepwise fashion, where the major 
steps correspond to changes in width. Aggregate growth 
may here be regarded as a quasi-2D process. 
Potential A leads to less severe barriers to increases in 
width, and thereby to (asymmetric) 2D growth. 
With potential C, there is no interaction at all ($\eij=0$) at   
hp and pp interfaces, which prevents the formation of
aggregates with more than two stacked sheets.   
In this case, aggregate growth becomes an effectively 
1D process.         

In our model, aggregates can be assigned a length and 
a width, whereas growth in a third dimension does not occur 
due to the interaction geometry. The length $l$ and width $w$ 
can be conveniently defined via the inertia tensor. Specifically, 
we define $l=\sqrt{12\lambda_1^2/m+1}$ 
and $w=\sqrt{12\lambda_2^2/m+1}$, 
where $\lambda_1\ge\lambda_2$ are eigenvalues of this 
tensor and $m$ is the number of peptides in the aggregate. 
This definition is such that a 
rectangular aggregate consisting of $w$ stacked sheets 
with $l$ peptides each ($l\ge w$) is assigned exactly 
length $l$ and width $w$.

\subsection{MC simulations}

In order to determine the thermodynamics of these systems, 
one needs simulations in which large aggregates form and 
dissolve many times, which is challenging to achieve even 
in a simple model. For our thermodynamic simulations, we 
therefore use a Swendsen-Wang--type cluster 
move~\cite{Swendsen:87} and a flat-histogram 
procedure.\cite{Berg:91,Hansmann:93,Wang:01a,Jonsson:11} 
  
The cluster move is based on a stochastic cluster construction 
scheme.\cite{Swendsen:87} The procedure is recursive and 
begins by picking a random first cluster member, $i$. Then, all peptides $j$
interacting with peptide $i$ ($\eij<0$) are identified and added to the 
cluster with probability $P_{ij}=1-\e^{\beta \eij}$, where 
$\beta$ is inverse temperature. 
This step is iterated until there are no more peptides 
to be tested for inclusion in the growing cluster. The resulting cluster 
is subject to a trial rigid-body translation or rotation, 
which is accepted whenever it does not cause any steric clashes. 
Unlike simpler cluster moves, this update can split and 
merge aggregates. The update fulfills detailed balance  
with respect to the canonical microstate 
distribution, $P_\nu\propto\e^{-\beta E_\nu}$.       
 
To further enhance the sampling, we use the multicanonical 
method,\cite{Berg:91,Hansmann:93,Wang:01a} 
which can be very useful for systems with multimodal 
energy landscapes. Our simulation procedure consists of
three steps.\cite{Jonsson:11} First,  we estimate the 
density of states, $g(E)$, by the Wang-Landau 
method~\cite{Wang:01a} (an early variant of which 
was proposed in Ref.~\citen{Engkvist:96}). Second, keeping this
estimate, $\tilde g(E)$, fixed, we simulate the ensemble 
$P_\nu\propto1/\tilde g(E_\nu)$, whose energy distribution 
is approximately flat.\cite{Berg:91} Finally, we calculate 
canonical averages via reweighting to the 
desired temperature.\cite{Ferrenberg:89} 
Throughout these simulations, we restrict the sampling 
to energies above a cutoff $\Emin$. This cutoff is taken
sufficiently high to avoid sampling of states containing 
unphysical cyclic aggregates, but sufficiently low to 
permit unbiased studies over the temperature  
range of interest. A more advanced, diffusion-optimized 
generalized-ensemble method was recently tested on 
this model.\cite{Tian:14} For our present purposes, the 
speed-up brought by the flat-histogram method suffices.     

In our flat-histogram simulations,  we use both single-peptide 
moves and the cluster move. The cluster move, as defined above, does 
not satisfy detailed balance with respect to   
$P_\nu\propto1/\tilde g(E_\nu)$. This can be easily
rectified by adding a Metropolis accept/reject step     
with the acceptance probability given by 
$\Pacc(\nu\to\nu^\prime)=
\min[1,g(E_\nu)\e^{-\beta E_\nu}/g(E_{\nu^\prime})\e^{-\beta E_{\nu^\prime}}]$.
Note that, in this context, $\beta$ is a tunable algorithm parameter, 
entering in the cluster construction, rather than a physical 
parameter. In our simulations, this parameter is chosen
in the vicinity of the inverse fibrillation temperature.  

In addition to the thermodynamic simulations, we perform 
relaxation simulations under fibril-favoring constant-temperature 
conditions. Here, motivated by experimental indications that 
amyloid growth occurs dominantly via monomer 
addition,\cite{Collins:04} we use single-peptide moves only;  
the elementary moves are translations by one 
lattice spacing and rotations of individual peptides.
Since there is no need to observe transitions back and forth 
between states with and without large aggregates, 
the systems can be much larger than in the 
thermodynamic simulations ($\sim$$10^5$ rather than 
$\sim$$10^2$ peptides).        

All statistical uncertainties quoted below are 1$\sigma$ errors.

\subsection{Semi-analytical approximation}

In this section, we present a semi-analytical method 
which can be used to estimate thermodynamic properties   
of the model for system sizes not amenable to direct simulation.
Let $\xi$ denote a certain configuration of $m_\xi$ peptides 
forming an aggregate, and let $N_\xi$ denote the number 
of such aggregates in the system. Treating the $N_\xi$'s 
as independent variables, the chemical potential for aggregates 
with configuration $\xi$ may be defined as 
$\mu_\xi \equiv \partial F / \partial N_\xi$, where  
$F(T,V,\lbrace N_\xi \rbrace)$ is a Helmholtz free 
energy. Imposing that 
the total number of peptides $\sum_\xi m_\xi N_\xi$ adds 
up to $N$, the function
\begin{equation}
\tilde{F} = F + \lambda \left( \sum_\xi m_\xi N_\xi -N \right),
\end{equation}
is minimized at equilibrium. After
eliminating the Lagrange multiplier $\lambda$, this leads 
to an equilibrium condition on the chemical potentials, namely
\begin{equation}
\mu_\xi = m_\xi \mu_1, \label{chemEq}
\end{equation}
where $\mu_1$ is the monomer chemical potential. 

With a simplified grand-canonical description, the partition
function for the set of aggregates with configuration $\xi$ is given by 
\begin{equation}
\mathcal{Z}_\xi = \sum_{N_\xi=0}^{\infty} \frac{\e^{\beta \mu_\xi N_\xi}}{N_\xi !} {Z^{(1)}_\xi}^{N_\xi} = \exp \left[ gV \e^{\beta (\mu_\xi - E_\xi)} \right],
\label{eq:z}\end{equation}
where $Z^{(1)}_\xi = gV \e^{-\beta E_\xi}$
is the single-aggregate partition function, $E_\xi$ is the 
internal energy, and $g=24$ 
is the number of possible spatial orientations of the aggregate. 
This description neglects interactions
between aggregates, but note that two adjacent $\xi$ 
aggregates correspond 
to one aggregate of some other type $\xi^\prime$. 
Eq.~\ref{eq:z} implies that the $N_\xi$ variable 
is Poisson distributed with mean
\begin{equation}
\langle N_\xi \rangle = 
\frac{1}{\beta} \pDer{ \log \mathcal{Z}_\xi}{\mu_\xi} =
gV \e^{\beta (\mu_1 m_\xi - E_\xi)}, \label{avNxi}
\end{equation}
where Eq.~\ref{chemEq} has been used. 
Given $\beta$, $V$ and $N$,
the monomer chemical potential 
can be determined by approximating 
$N_\xi \approx \langle N_\xi \rangle$ and solving
\begin{equation}
N = \sum_\xi m_\xi N_\xi,	\label{pepCons}
\end{equation}
for $\mu_1$. Knowing $\mu_1$, one can obtain 
the $N_\xi$'s from Eq.~\ref{avNxi} and compute 
the total energy as
\begin{equation}
E = \sum_\xi E_\xi N_\xi\,.
\label{eav}
\end{equation}
The fibrillation temperature $\Tf$ may be
defined as the maximum of the heat capacity, and can therefore
be estimated by numerically differentiating Eq.~\ref{eav}. 
This procedure for computing $\Tf$ is fast and can be repeated for many  
different concentrations $c=N/V$, once the relevant aggregates 
$\xi$ and their energies $E_\xi$ are specified. Hence, it can be 
used to estimate the state of the system as a function of both 
$T$ and $c$. 

The above scheme has similarities to the approach of Oosawa and 
Kasai,\cite{Oosawa:62} but has here been derived for more   
general choices of included aggregates.   
When applying it to the present model, 
we only consider rectangular 
aggregates with an energetically optimal internal 
organization. The generic index $\xi$ can therefore 
be replaced by the aggregate length $l$ and 
width $w$. This set of configurations 
turns out to be sufficient to obtain quite accurate estimates 
of $\Tf $. To respect the finite size of the systems,  
we limit the sums over $l$ and $w$ to $l\le l_{\max} = \min(N,L)$ 
and $w\le w_{\max}(l) = \min \left[ L, \mathrm{floor}\left(N/l\right) \right]$.
For the potential C, which leads to aggregates 
with at most two proper sheets, we set
$w_{\max}(l) = \min \left[ 2, \mathrm{floor}(N/l)\right]$.  

The energy of an aggregate with length $l$ and width $w$, 
$E_{lw}$, is a sum of intra- and intersheet contributions. In
the minimum energy configuration, the intrasheet energy is 
$-(1+\apa) (l-1)w$, corresponding to a parallel organization 
of the $l$ peptides, whereas the intersheet energy depends on the 
parameters $\bhh$, $\bhp$ and $\bpp$. For the potentials 
A, B and C (Table~\ref{tab:b}), the respective minimum total 
energies are given by
\begin{eqnarray}
E^{\mathrm{(A)}}_{lw} &=& -(1+\apa)(l-1)w - \frac{3l(w-1)}{2}\label{EA}\\
E_{lw} ^{(\mathrm{B})} &=& 
-(1+\apa)(l-1)w - \frac{3l(w-1)}{2}-\frac{(1+(-1)^w)l}{4}\label{EB}\\
E_{lw} ^{(\mathrm{C})}&=&-(1+\apa)(l-1)w - 2 l(w-1)\qquad (w=1,2)
\end{eqnarray}
With potential B, the minimal energy is achieved when  
both outer surfaces are entirely polar, which maximizes the  
number of favorable hh contacts. Likewise, with potential C, 
the energy of a two-sheet aggregate is minimal if the entire
interface is of hh type.

\section{\label{sec:results}Results and discussion}

We study both equilibrium and relaxation properties of  
the above model for the three choices of intersheet 
interaction parameters listed in Table~\ref{tab:b}, which 
correspond to 2D, quasi-2D and 1D growth and are 
referred to as A, B and C, respectively. 
The intrasheet interactions, which are stronger, 
stay the same in all three cases.   
  
\subsection{Equilibrium properties}

The equilibrium properties of the model are investigated by 
using MC simulations and the semi-analytical approximation 
described in Sec.~\ref{sec:methods}. We first present the MC results which, 
unless otherwise stated, are obtained using $N=256$ 
and $L=64$, corresponding to a 
concentration of $c=0.977\times10^{-3}$ per unit volume.     

\begin{figure}[t]
\centering
  \includegraphics[width=8cm]{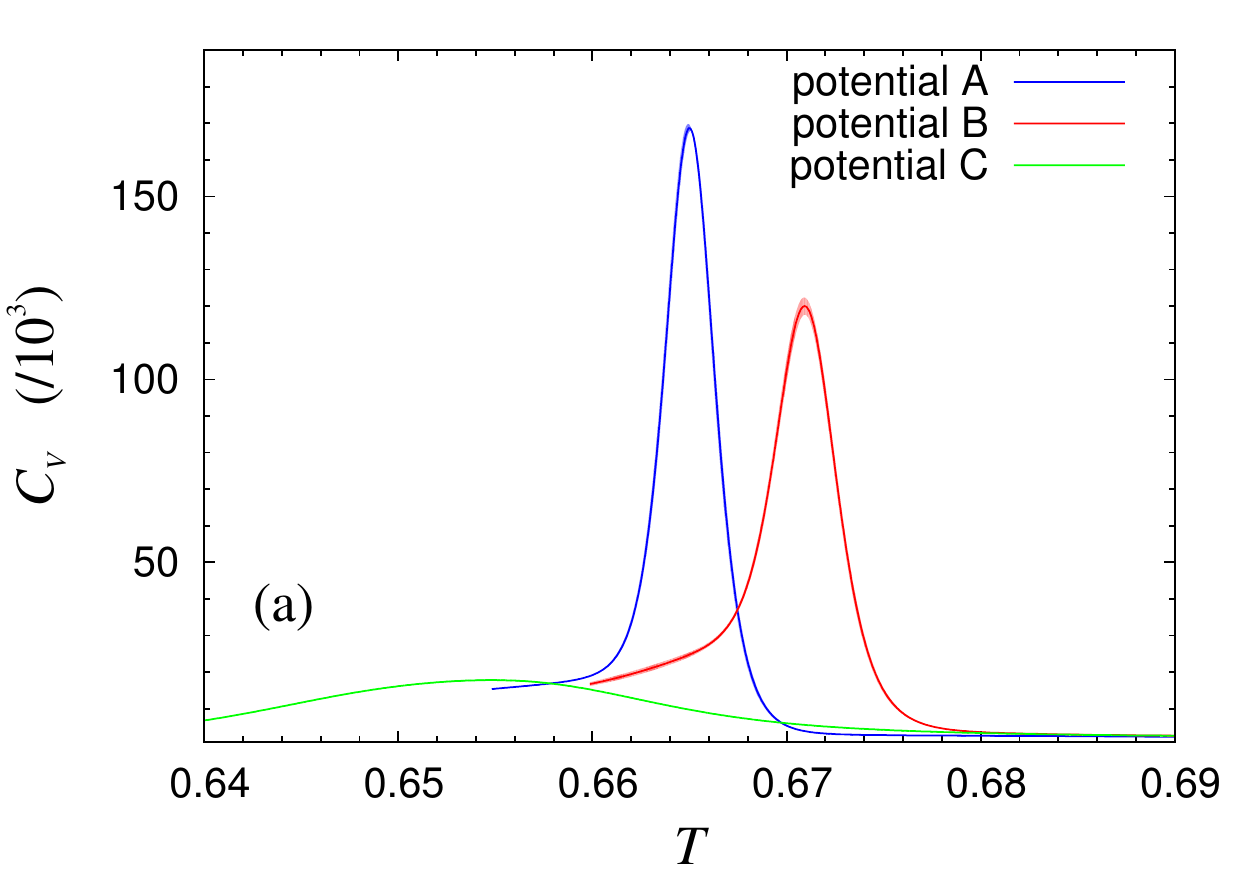}
  \includegraphics[width=8cm]{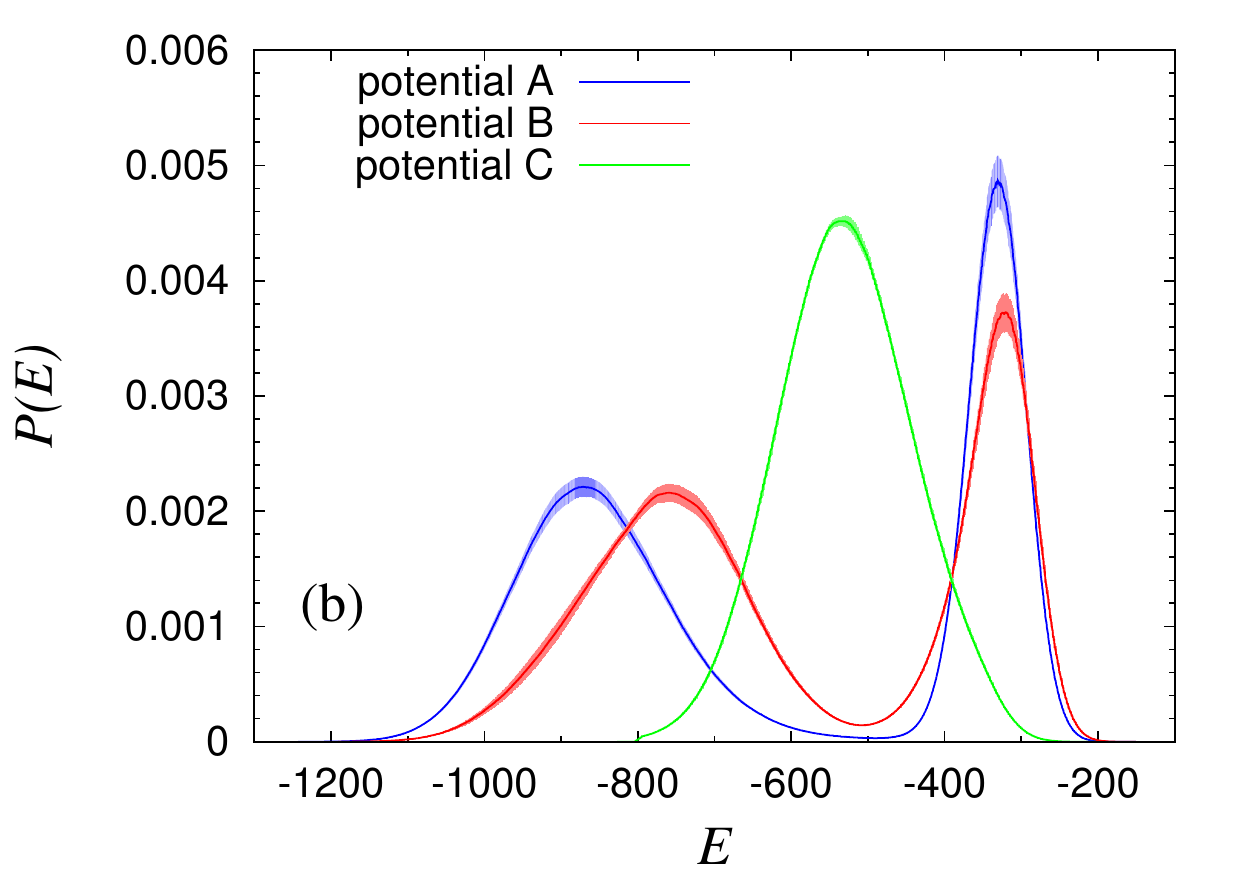}
  \caption{(a) Temperature-dependence of the heat capacity, $\Cv$,
  from MC simulations with $N=256$ and $L=64$ for the 
  potentials A, B and C (Table~\ref{tab:b}). The shaded 
  bands indicate statistical uncertainty. Our flat-histogram  
  simulations sample energies $E>E_{\min}$, where $E_{\min}$ is a
  cutoff. Data are shown only at temperatures that are sufficiently   
  high for the effects of this cutoff to be negligible.
  (b) Distribution of energy, $P(E)$, at $T=\Tf $, 
  for the same three systems. Due to short-scale irregularities 
  in the density of states, a moving average is used 
  (window size $\Delta E=10$). 
  \label{fig:cv}}
\end{figure}

All three systems contain large fibril-like aggregates at low $T$, 
while being disordered at high $T$. The onset of fibril 
formation is accompanied by a peak in the 
heat capacity $\Cv=(\ev{E^2}-\ev{E}^2)/T^2$ (Fig.~\ref{fig:cv}a).  
The fibrillation temperature, $\Tf$, may therefore be defined 
as the maximum of $\Cv$, and is found to be given by    
$\Tf ^\text{(A)}=0.66500\pm0.00007$,  
$\Tf ^\text{(B)}=0.67093\pm0.00007$ and 
$\Tf ^\text{(C)}=0.6548\pm0.0002$ 
for systems A, B and C, respectively.  

While $\Tf $ is thus roughly similar for all three systems, there
are large differences in the height of the $\Cv$ peak (Fig.~\ref{fig:cv}a).
This fact reflects a fundamental difference between systems
A and B, on one hand, and system C, on the other hand, 
as can be seen from the probability distribution of the total 
system energy $E$ (Fig.~\ref{fig:cv}b). For systems A and B, 
with a pronounced peak in $\Cv$, the energy distribution 
is clearly bimodal, showing that these systems can exist in 
two distinct types of states at $T=\Tf$. The difference
between the two potentials shows up in the location of the 
low-energy peak. For system C, the $\Cv$ peak 
is broader and lower. In this system, the onset of fibril 
formation is smooth. As the temperature is reduced, 
the energy distribution slides toward lower values while 
retaining a unimodal shape.  

\begin{figure}[t]
\centering
  \includegraphics[width=8cm]{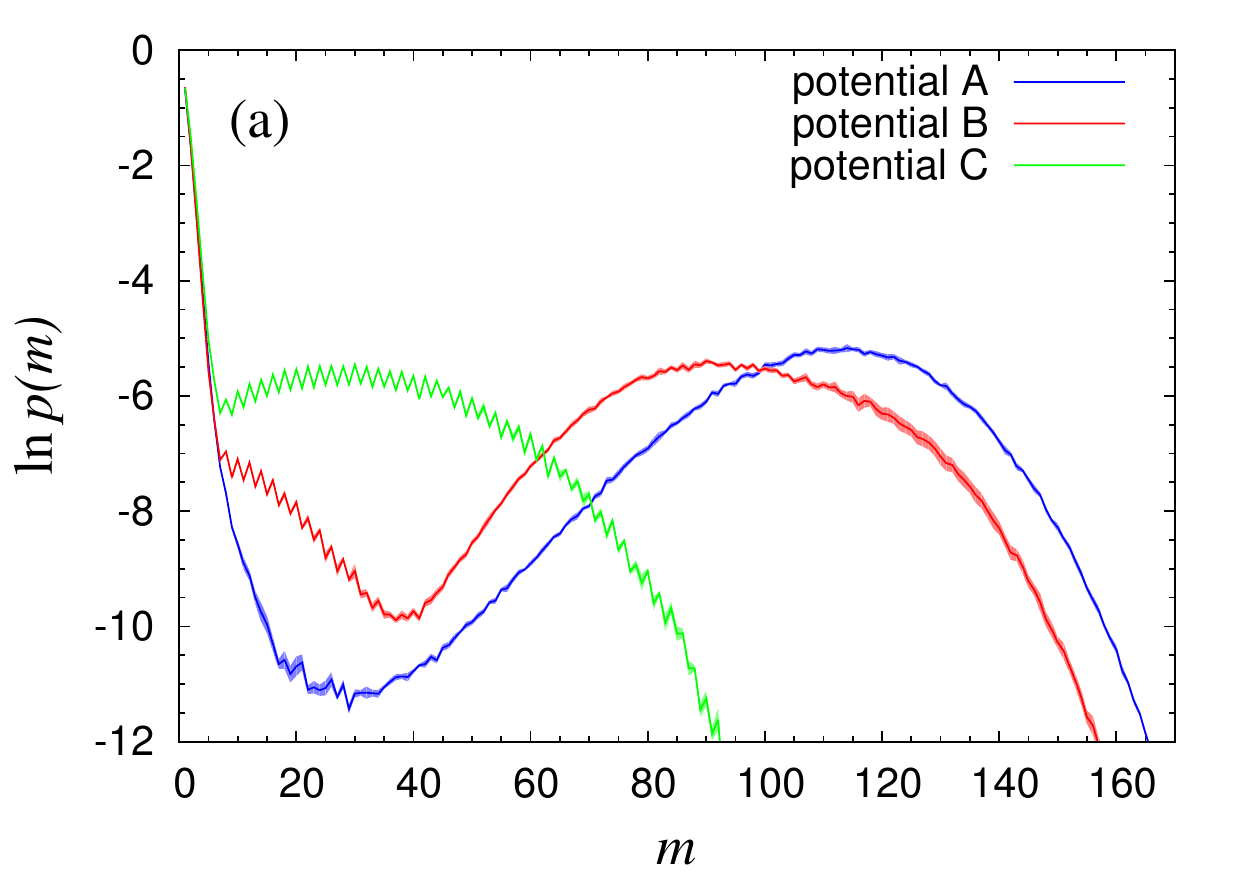}
  \includegraphics[width=8cm]{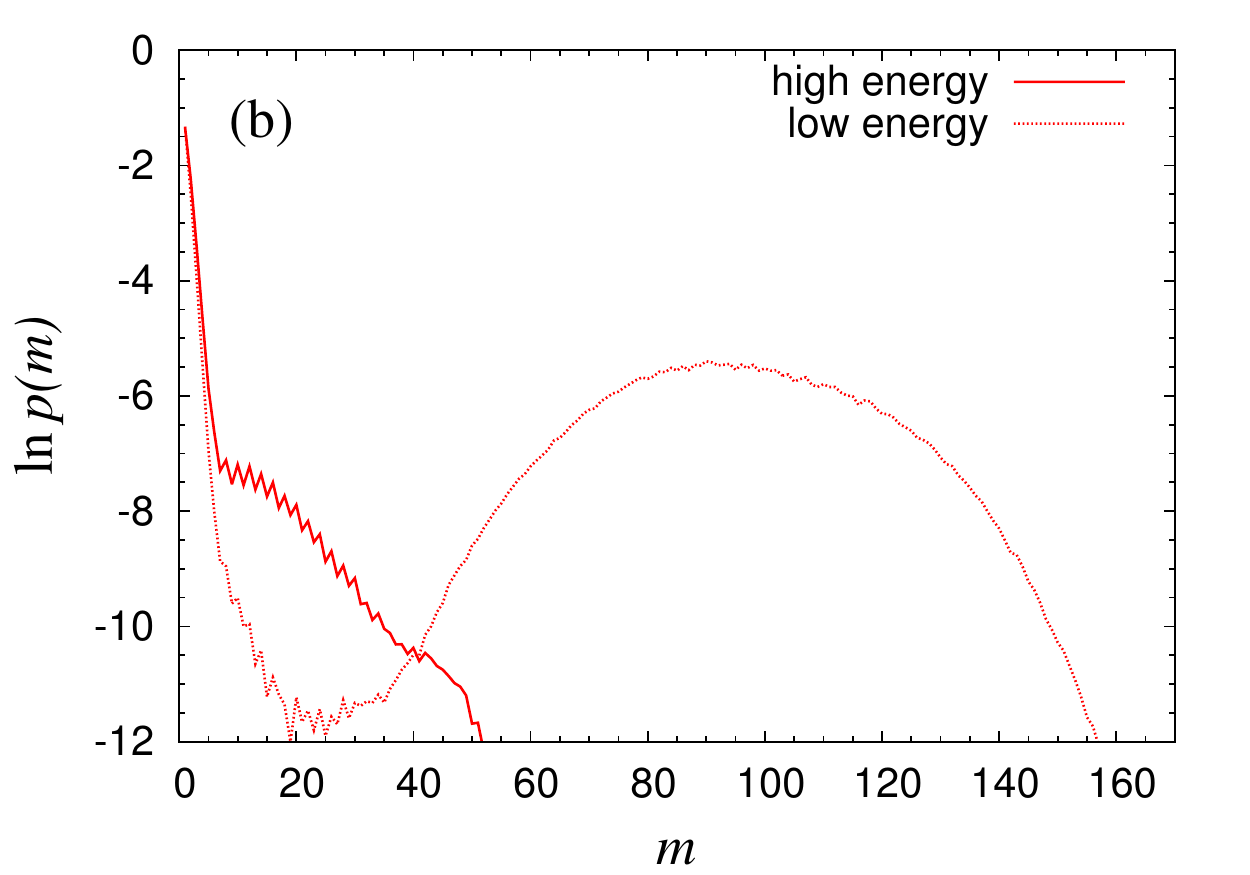}
  \caption{(a) Mass fraction of aggregates with size $m$, $p(m)$, 
  against $m$ at $T=\Tf $, as obtained 
  from MC simulations with $N=256$ and $L=64$
  for the potentials A, B and C (Table~\ref{tab:b}).  
  The shaded bands indicate statistical uncertainty. 
  The saw-tooth-like behavior that occurs for systems 
  B and C is due to even-odd effects for 
  two-layered aggregates. (b) Decomposition of 
  $p(m)$ for system B into contributions from 
  low- and high-energy states, respectively 
  (compare Fig.~\ref{fig:cv}b). Each 
  configuration in the simulated ensemble is classified as 
  either low energy ($E<-500$) or high energy ($E\ge-500$). 
  \label{fig:size}}
\end{figure}

The behavior of the systems at the fibrillation 
temperature can be further characterized in terms of the 
aggregate-size distribution, $p(m)$, which gives the 
probability for a random peptide to be part of an aggregate 
with size $m$ (Fig.~\ref{fig:size}a). 
For systems A and B, $p(m)$ is bimodal at $T=\Tf $. 
Hence, whereas both small and large aggregates occur 
in these systems, there is a range of suppressed  
intermediate sizes. Above, it was seen that the energy
distribution is bimodal as well (Fig.~\ref{fig:cv}b). States 
belonging to the low-energy peak contain both small and 
large aggregates and contribute, therefore, to both peaks 
in $p(m)$, whereas high-energy states are dominated 
by small aggregates. Fig.~\ref{fig:size}b shows 
the contributions to $p(m)$ from low- and high-energy 
states in system B, which indeed are bi- and unimodal,
respectively. Also worth noting in this figure is that the
amount of aggregates with size between $m\approx7$ and
$m\approx35$ tends to be much smaller in low-energy
states than in high-energy states. Hence, the appearance 
of large aggregates in low-energy states occurs, at least in part, 
at the expense of these mid-size ones. For system
C, the aggregate-size distribution $p(m)$ is fundamentally 
different (Fig.~\ref{fig:size}a). In this system, there is no 
intermediate range of suppressed sizes $m$, and therefore 
no clear division into either small or large species. 

\begin{figure}[t]
\centering
  \includegraphics[width=8cm]{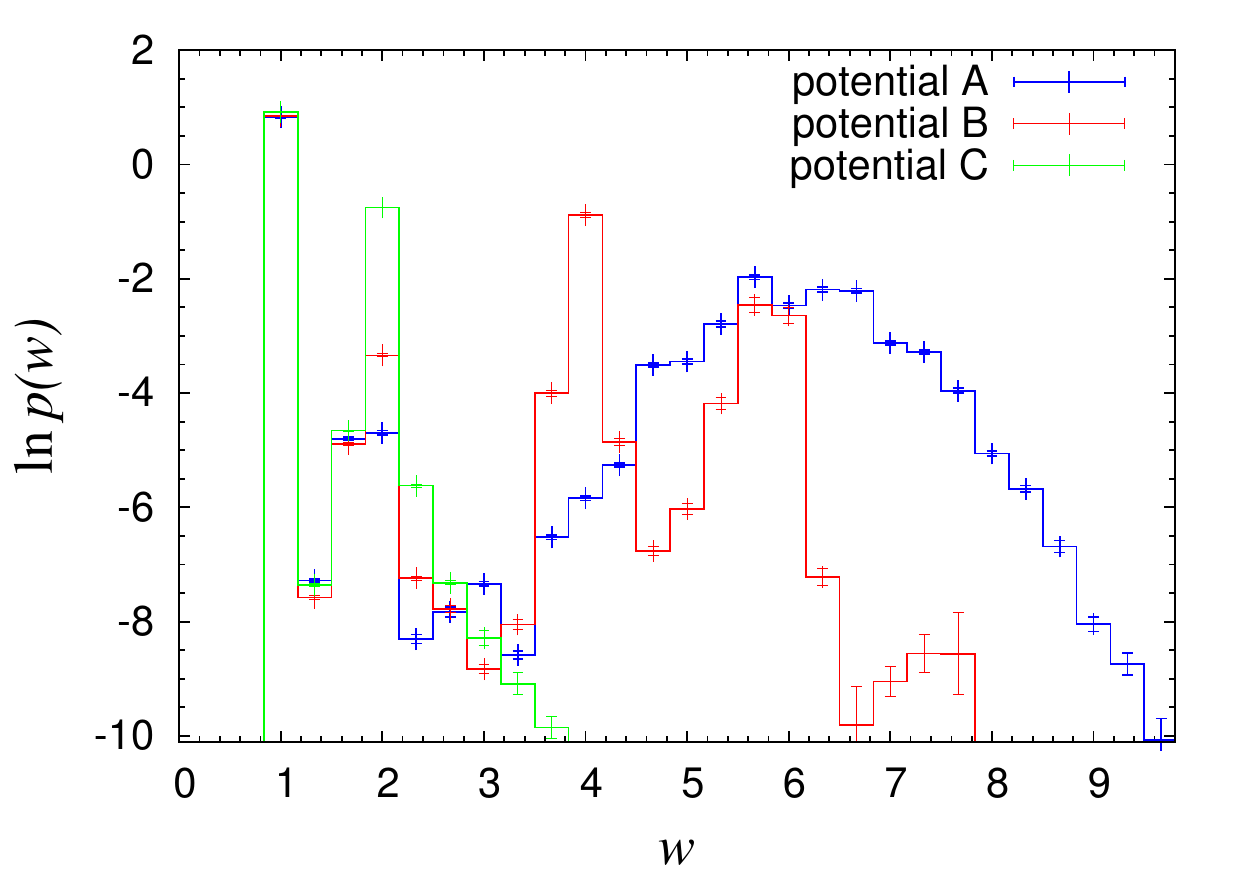}
  \caption{Mass fraction of aggregates with width $w$, $p(w)$, 
  against $w$ at $T=\Tf $, as obtained 
  from MC simulations with $N=256$ and $L=64$
  for the potentials A, B and C (Table~\ref{tab:b}).    
  \label{fig:width}}
\end{figure}

The intersheet interactions directly influence the width 
of the aggregates, $w$ (see Sec.~\ref{sec:methods}). Fig.~\ref{fig:width}
shows the mass-weighted distribution of $w$, $p(w)$, at $T=\Tf$ for 
our three systems.  As expected, $p(w)$ decays rapidly  
beyond $w=2$ for potential C, whereas potentials A and B 
permit the formation of wider aggregates (Fig.~\ref{fig:width}). 
The data also confirm that the asymmetric intersheet interactions 
of potential B indeed favor even-layered aggregates  
over odd-layered ones.
 
\begin{figure}[t]
\centering
  \includegraphics[width=8cm]{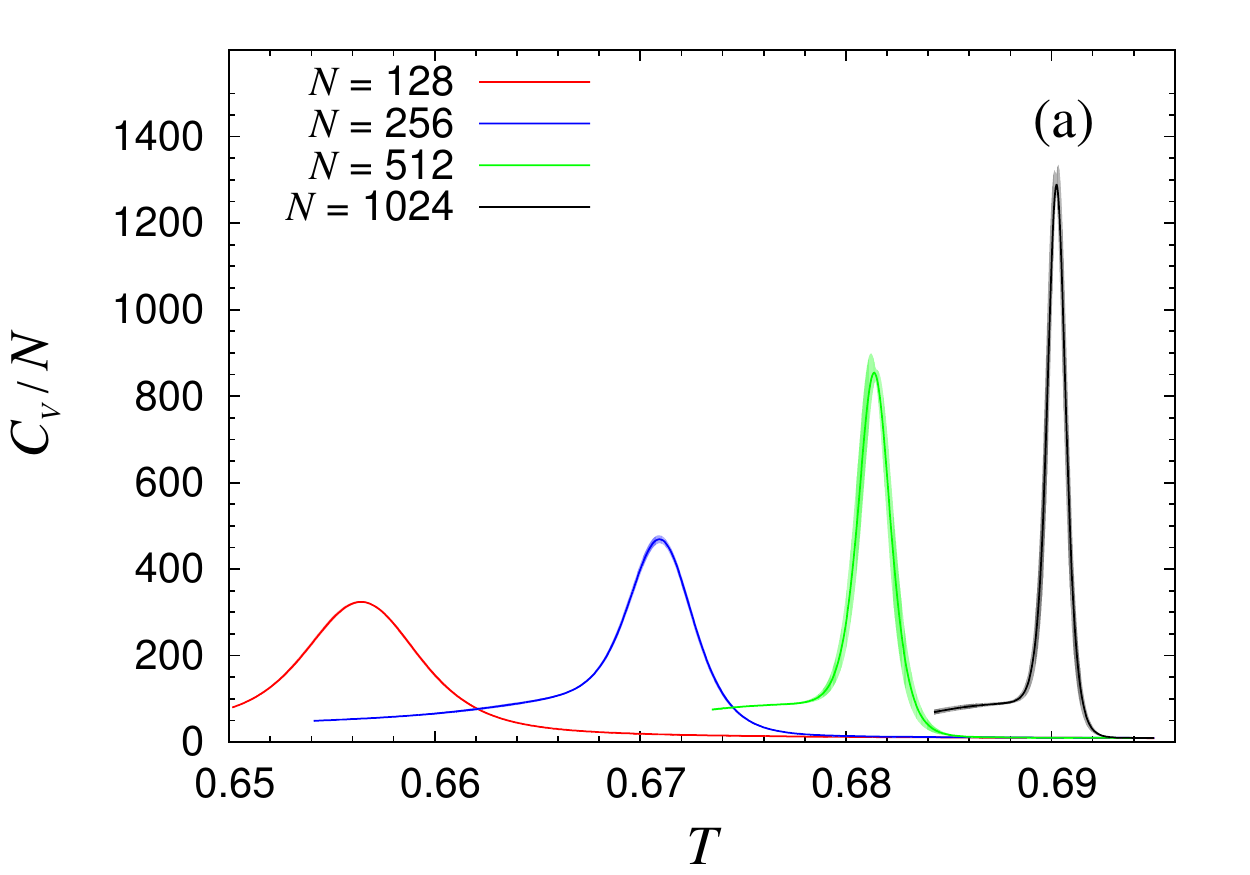}
  \includegraphics[width=8cm]{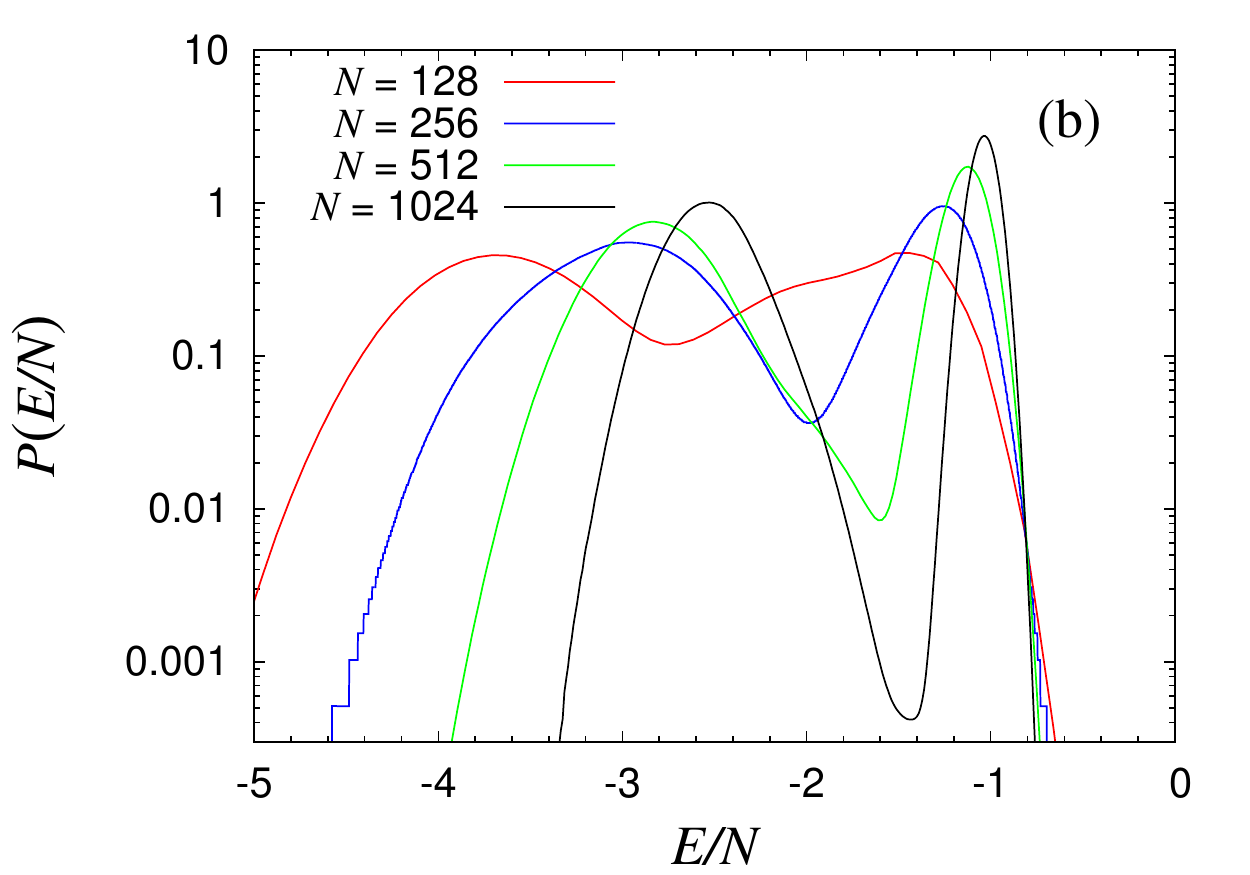}
  \caption{Finite-size scaling for system B (Table~\ref{tab:b}),  
  at fixed concentration. The data are from MC simulations 
  with $N=128$, 256, 512 and 1028 ($L=51$, 64, 81 and 102).     
  (a)  Temperature dependence of the specific heat, $\Cv/N$.
  The shaded bands indicate statistical uncertainty. 
  (b)  Probability distribution of the energy density $E/N$ at the  
  specific-heat maximum $\Tf$.    
  \label{fig:fss}}
\end{figure}

The above discussion focused on results obtained
using $N=256$ and $L=64$. To better understand the sharp onset
of fibril formation in systems A and B, additional simulations were 
performed for a few different $N$, keeping the concentration
approximately constant. Fig.~\ref{fig:fss}a shows the specific heat, 
$\Cv/N$, of system B for $N=128$, 256, 512 and 1024. 
As $N$ increases, the peak in $\Cv/N$
gets sharper. However, the height of the peak increases more slowly
than the linear growth expected at a first-order phase transition 
with a non-zero specific latent heat. Indeed, the latent heat, or energy 
gap, does not scale linearly with $N$ (Fig.~\ref{fig:fss}b). Still, 
the gap grows sufficiently fast (faster than $N^{1/2}$) 
for the bimodality of the energy distribution to become more and more 
pronounced with increasing $N$ (Fig.~\ref{fig:fss}b).       
Therefore, fibril formation sets in at a 
first-order-like transition, where distinct states
coexist.  Similar analyses 
were performed for potentials A and C, using $N=128$, 
256 and 512. The results obtained with potential A are 
qualitatively similar to those just described for potential B. For potential C, 
$\Cvm/N$ does not grow with $N$, thus confirming the conclusion 
that, in this case, the onset of fibril formation 
represents a crossover rather than a sharp transition.

Our systems resemble a lattice gas at fixed particle number, albeit   
with asymmetric interactions. For finite-volume liquid-vapor systems 
at phase coexistence, the formation of droplets due to a fixed particle excess
above the ambient gas concentration has been extensively 
investigated,\cite{Binder:80,Furukawa:82,Biskup:02,Biskup:03,Neuhaus:03,
Biskup:04,MacDowell:04,Nussbaumer:06,Nussbaumer:10,Bauer:10,Nogawa:11} 
often by mapping to the Ising model at fixed magnetization. A sharp transition 
has been shown to occur, below which the particle excess can be accommodated
by gas-phase fluctuations. At the transition point, a large droplet 
appears, whereas intermediate-size droplets remain strongly suppressed.
The volume of the droplet and the latent heat scale as $V^{3/4}$.  
To accurately determine the corresponding scaling behavior for our systems A and B, data over
a wider range of system sizes would be required. However, the scaling of
the latent heat does seem to be faster than $V^{1/2}$ and slower than $V$
(Fig.~\ref{fig:fss}b). 

\begin{figure}[t]
\centering
  \includegraphics[width=8cm]{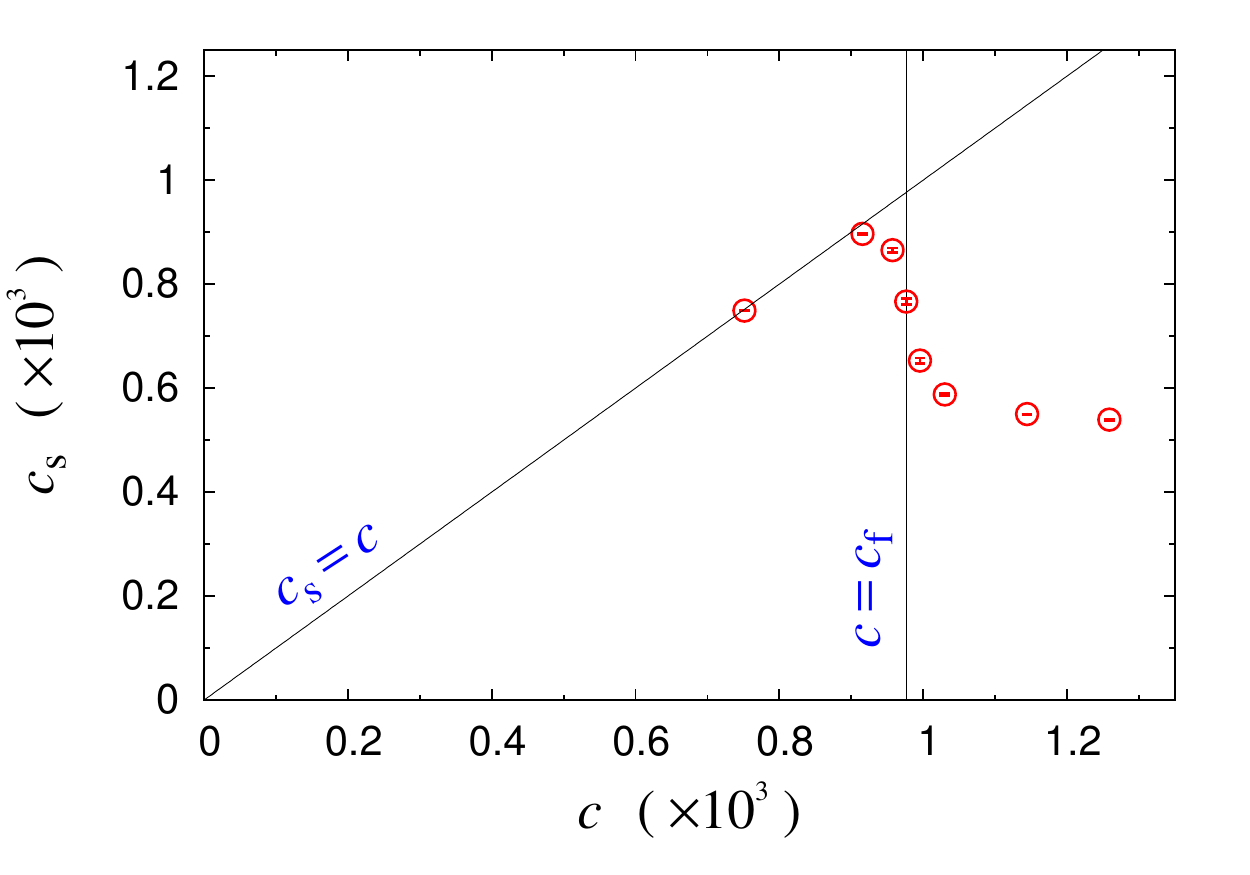}
  \caption{
  Concentration of free peptides, $\cs$, against total concentration, $c$, 
  as obtained by MC simulations of system B  (Table~\ref{tab:b}) for
  $T=0.67093$, $L=64$ and different $N$. At this temperature, fibrillation   
  sets in at $\cf\approx256/64^3$. Our definition of free peptides includes 
  monomers and aggregates with size $m\le 6$, so $\cs=c\times\sum_{m=1}^6 p(m)$.
  \label{fig:ms_vs_c}}
\end{figure}

At the droplet condensation transition, the gas concentration 
drops by an amount that scales as $V^{-1/4}$.\cite{Biskup:02,MacDowell:04} 
In our systems A and B, at the threshold concentration for fibril formation, $\cf(T)$,  
a similar drop occurs in the concentration of free peptides, $\cs$, 
as is illustrated in Fig.~\ref{fig:ms_vs_c} 
by data obtained with potential B for $L=64$ and different $N$. 
To test the scaling with system size, this drop, $\Delta \cs$,   
was computed at the heat-capacity maxima of Fig.~\ref{fig:fss}a, 
for system B and four different $V$. These $\Delta \cs$ values 
vary roughly as $V^{-1/4}$ with $V$  
($\Delta\cs\times V^{1/4}=0.0089$, 0.0079, 0.0085 and 0.0086 
for $N=128$, 256, 512 and 1024, respectively).   
    
\begin{figure}[t]
\centering
  \includegraphics[width=8cm]{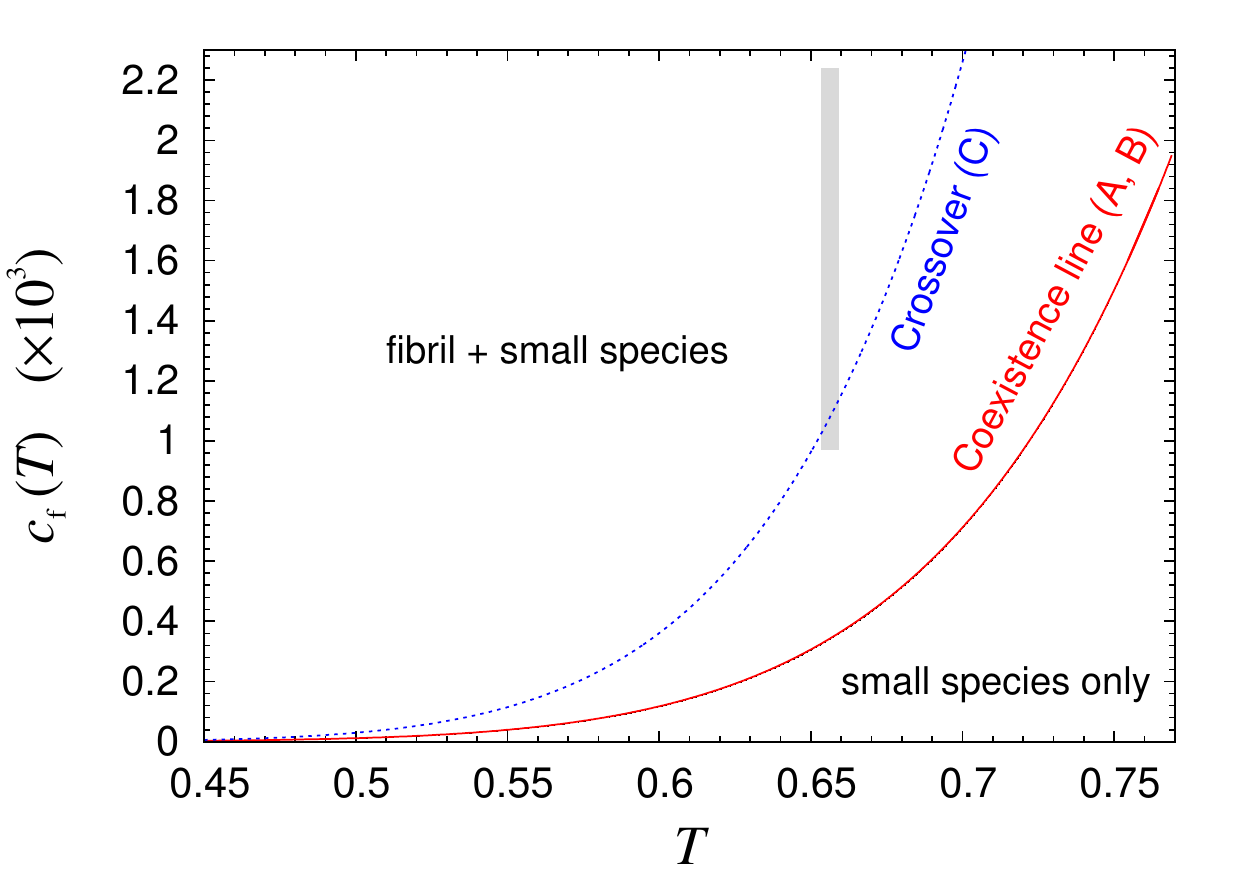}
  \caption{
  Temperature-dependence of the threshold concentration, $\cf(T)$, for systems 
  A, B and C (Table \ref{tab:b}), as obtained by the semi-analytical 
  approximation (Sec.~\ref{sec:methods}) for $L=512$. For system C, $\cf(T)$
  represents a crossover rather than a sharp transition. The 
  grey vertical bar indicates the $T,c$ interval studied in our relaxation simulations.
  \label{fig:phasediagram}}
\end{figure}

The results presented so far were obtained by 
MC simulations, which are bias-free but time-consuming.  
To be able to study larger systems, the approximate but much 
faster semi-analytical approach (Sec.~\ref{sec:methods}) 
is used.  For $L=64$ and $N=256$, this method provides
estimates of the fibrillation temperature  
($\Tf ^{\text{(A)}} = 0.6783$, $\Tf ^{\text{(B)}} = 0.6866$, 
$\Tf ^{\text{(C)}} = 0.6509$) that
agree to within $\sim$1\% with the MC results, 
and the ordering $\Tf ^{\text{(C)}}<\Tf ^{\text{(A)}}<\Tf ^{\text{(B)}}$ 
is correct. Having seen this agreement, the method is 
applied to estimate the threshold concentration, $\cf(T)$, as a function 
of temperature  for the box size used in the relaxation 
simulations below, that is $L=512$. To this end, the 
fibrillation temperature is calculated for a large set of   
concentrations in the range $1.0\times10^{-6}<c<2.3\times10^{-3}$. 
The resulting estimates of $\cf(T)$ are shown in 
Fig.~\ref{fig:phasediagram}. The curves for systems 
A and B, with closely related energies (Eqs.~\ref{EA},\ref{EB}), agree to 
within $\sim$1\% at $T=0.65$ and become even more 
similar at higher $T$. For system C,  our method estimates a 
higher $\cf(T)$. As discussed above, in this system,  
$\cf(T)$ represents a crossover rather than a sharp transition. 
The same scheme also provides an estimate of the heat 
capacity.  It predicts $\Cv$ to vary smoothly with $T$ in
system C but that a jump occurs at $T=\Tf$ in systems A 
and B, all of which match well with our earlier conclusions 
based on MC data for smaller systems.  

\subsection{Relaxation simulations} 

Having located the threshold concentration $\cf(T)$ for fibril 
formation, we next study the relaxation of the systems in 
constant-temperature MC simulations with $c>\cf(T)$, starting 
from random initial states. Assuming fibril growth to occur 
through monomer addition,\cite{Collins:04} the simulations are 
performed using single-peptide moves only. The parameters 
$T=0.6535$ and $L=512$ are the same in all these calculations, 
whereas $N$ varies between $2^{17}=131,072$ and 300,000. The 
corresponding $c$ interval is indicated in Fig.~\ref{fig:phasediagram}. 
To assess statistical uncertainties, a set of eight independent runs 
is generated for each choice of concentration and potential.

\begin{figure}[t]
  \mbox{
  \includegraphics[width=6.5cm]{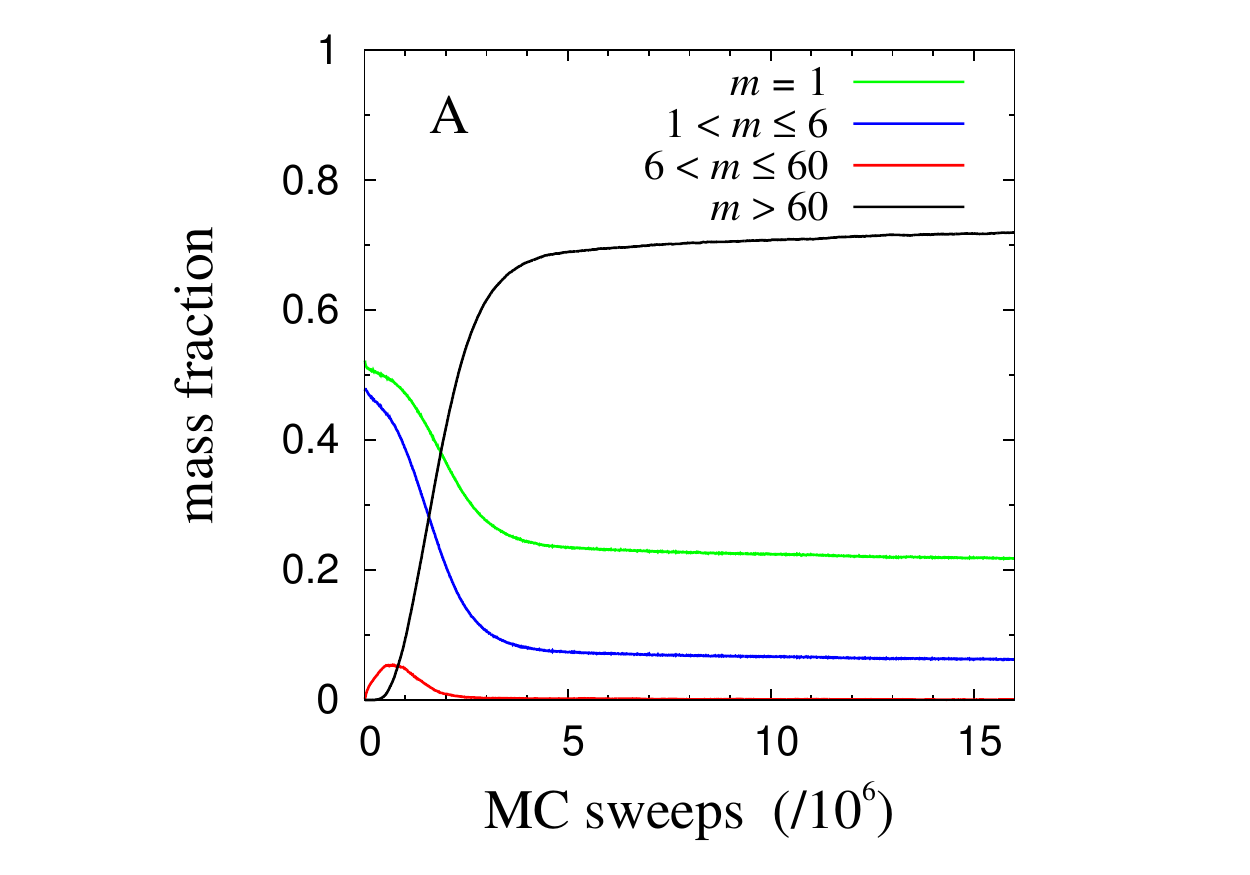}
  \hspace{-1.5cm}
  \includegraphics[width=6.5cm]{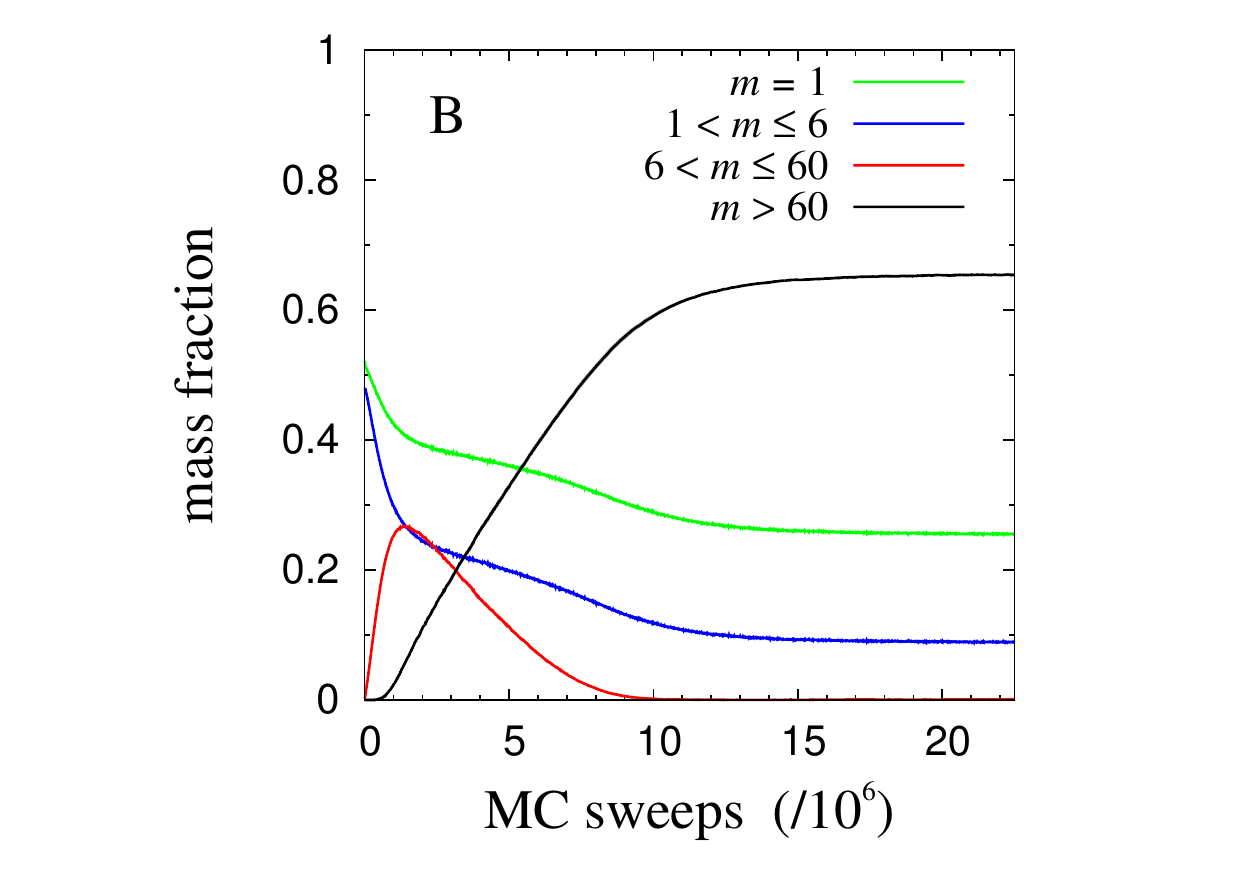}
  \hspace{-1.5cm}
  \includegraphics[width=6.5cm]{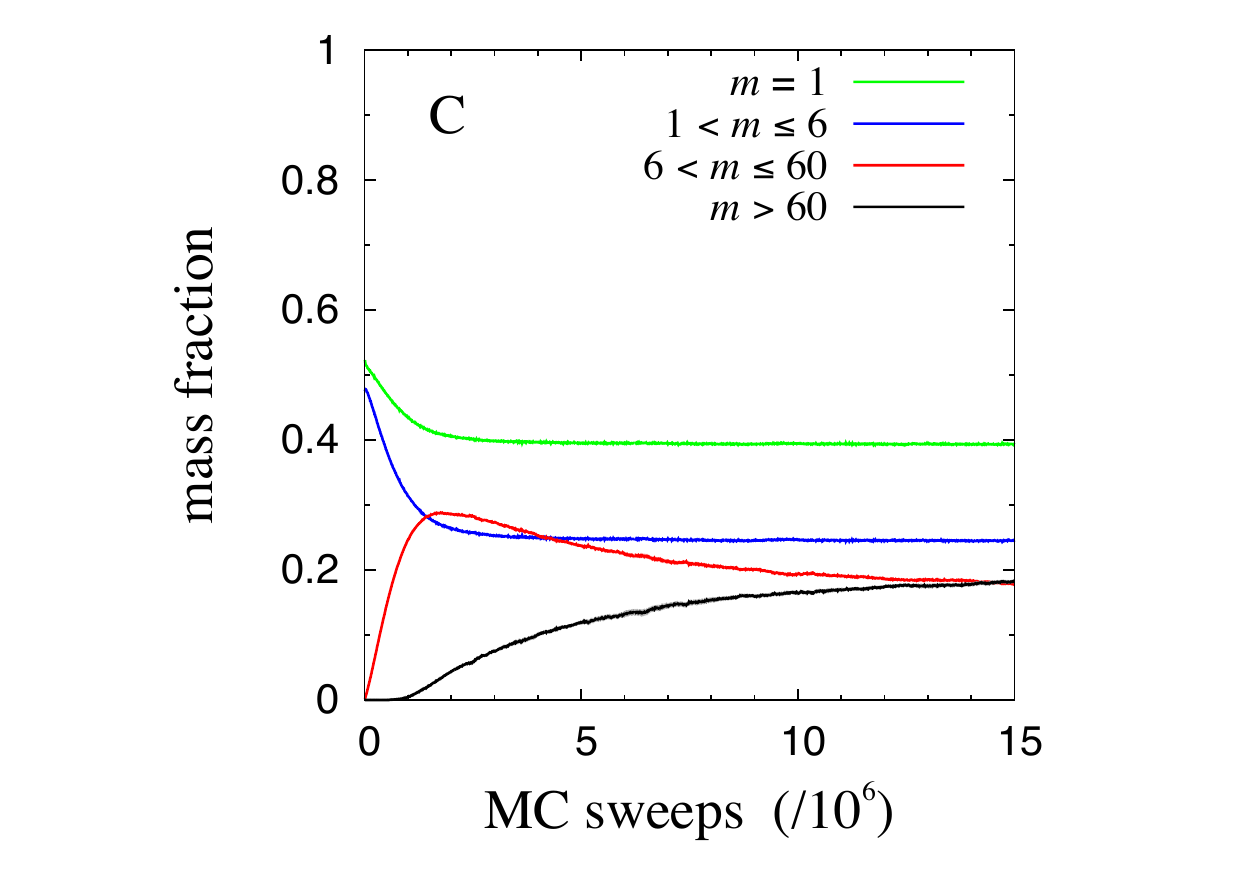}
  }
  \caption{
  MC time evolution of the mass fractions of monomers ($m=1$),
  small aggregates ($1<m\le 6$), mid-size aggregates ($6<m\le 60$)
  and large aggregates ($m>60$) in simulations with the 
  potentials A, B and C (Table~\ref{tab:b}), using $T=0.6535$, $L=512$ and 
  $N=170,000$. At large times, 
  small and large species dominate in systems A and B, 
  whereas mid-size aggregates remain present in system C. 
  Each MC sweep consists of $N$ single-peptide moves.
  The data represent averages over eight independent runs.
  The statistical uncertainties, indicated by shaded bands, are 
  barely visible.
  \label{fig:massdist}}
\end{figure}

\begin{figure}[t]
\centering
  \mbox{
  \includegraphics[width=8.5cm]{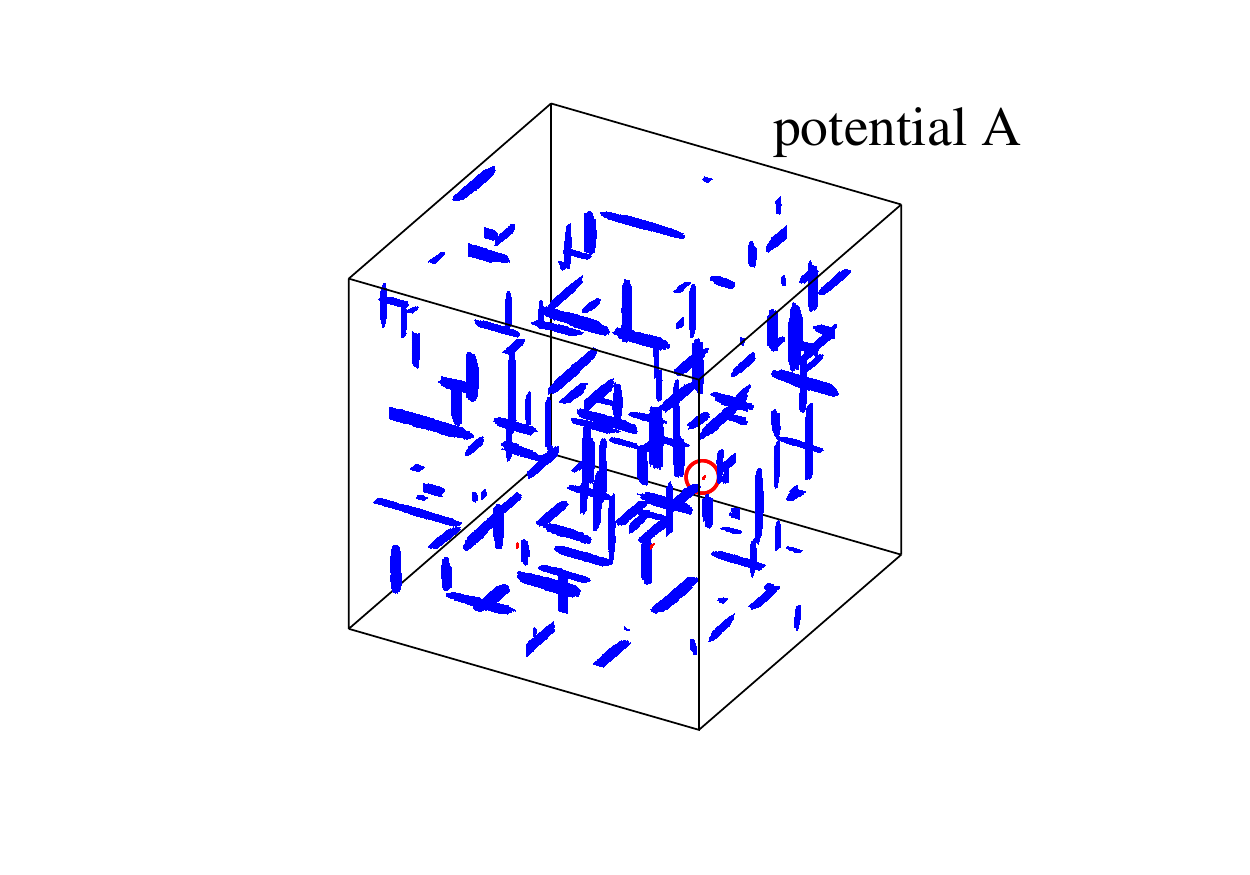}
  \hspace{-3cm}
  \includegraphics[width=8.5cm]{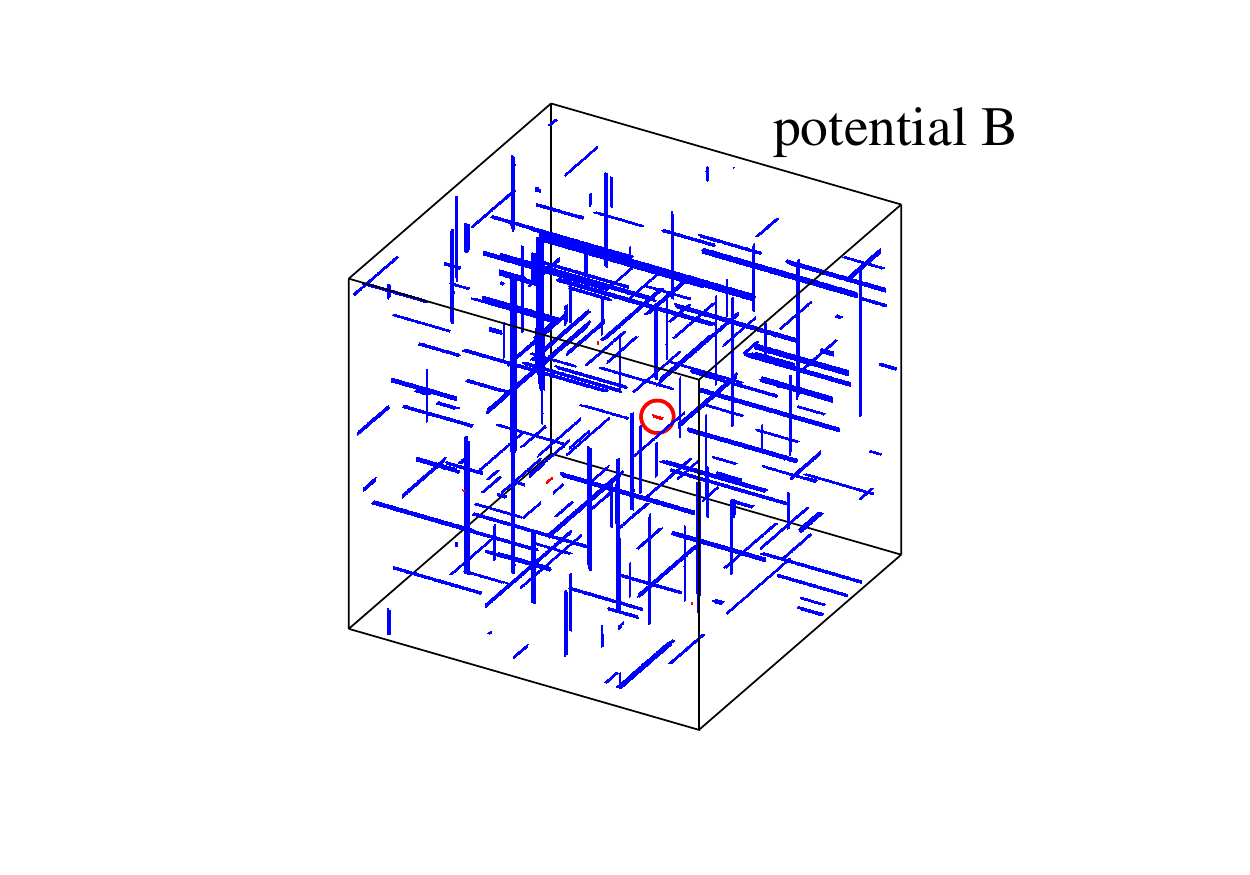}
  }
  \caption{
  Final configurations from relaxation simulations with potentials A and B 
  (Table~\ref{tab:b}), using $T=0.6535$, $L=512$ and $N=170,000$.   
  Large ($m>60$) and mid-size ($6<m\le60)$ aggregates are shown in 
  blue and red, respectively. For clarity, small species ($m\le6$) are not 
  shown. At this stage, the mid-size aggregates have almost disappeared 
  (compare Fig.~\ref{fig:massdist}). Red circles indicate one of $\le$5
  such aggregates in each configuration. Large aggregates tend to be 
  shorter and wider with potential A than they are with potential B.   
  \label{fig:snapshots}}
\end{figure}

Fig.~\ref{fig:massdist} 
illustrates how aggregation proceeds 
with potentials A, B and C, by showing the evolution of the respective 
mass fractions of (i) monomers, (ii) small aggregates 
with $1<m\le 6$ peptides, (iii) mid-size aggregates with $6<m\le60$,
and (iv) large aggregates with $m>60$. The number of peptides 
is the same in all three cases ($N=170,000$).
The monomer fraction is close to unity in the random initial 
states, but roughly a factor 2 smaller already at the time of the first 
measurement, due to rapid equilibration between monomers and
small aggregates. After this point, the amounts of monomers and small 
aggregates decrease monotonically toward apparent steady-state levels.   
The fate of the mid-size aggregates depends on the potential. 
In systems A and B, these aggregates are transient species. 
Examples of final configurations from the simulations of these 
systems can be found in Fig.~\ref{fig:snapshots}, both of which
contain many large fibril-like aggregates but only very few 
mid-size ones. In system C, there is, by contrast, a 
non-negligible amount of mid-size aggregates still present in 
the apparent steady-state regime.

The apparent steady-state regimes in these simulations need 
not correspond to thermodynamic equilibrium states. In fact,  it is likely that 
the true equilibrium states of the systems shown in Fig.~\ref{fig:snapshots} 
contain only one very large aggregate accompanied by surrounding 
small species, as observed in our equilibrium simulations of smaller 
systems. However, due to the very slow dynamics of large aggregates,
the states shown in Fig.~\ref{fig:snapshots} are effectively frozen on the 
timescales of our simulations. 

Finally, we also study how the overall rate of fibril formation scales with 
concentration in our relaxation simulations, focusing on systems A and B.
Here, an aggregate is taken to be a fibril if its width $w$ exceeds 
3.5, because thinner aggregates are unstable. This definition is
somewhat arbitrary, but ambiguous assemblies close to the cutoff 
in width are transient species that essentially disappear as 
aggregation proceeds.

\begin{figure}[t]
  \begin{center}
  \includegraphics[width=8cm]{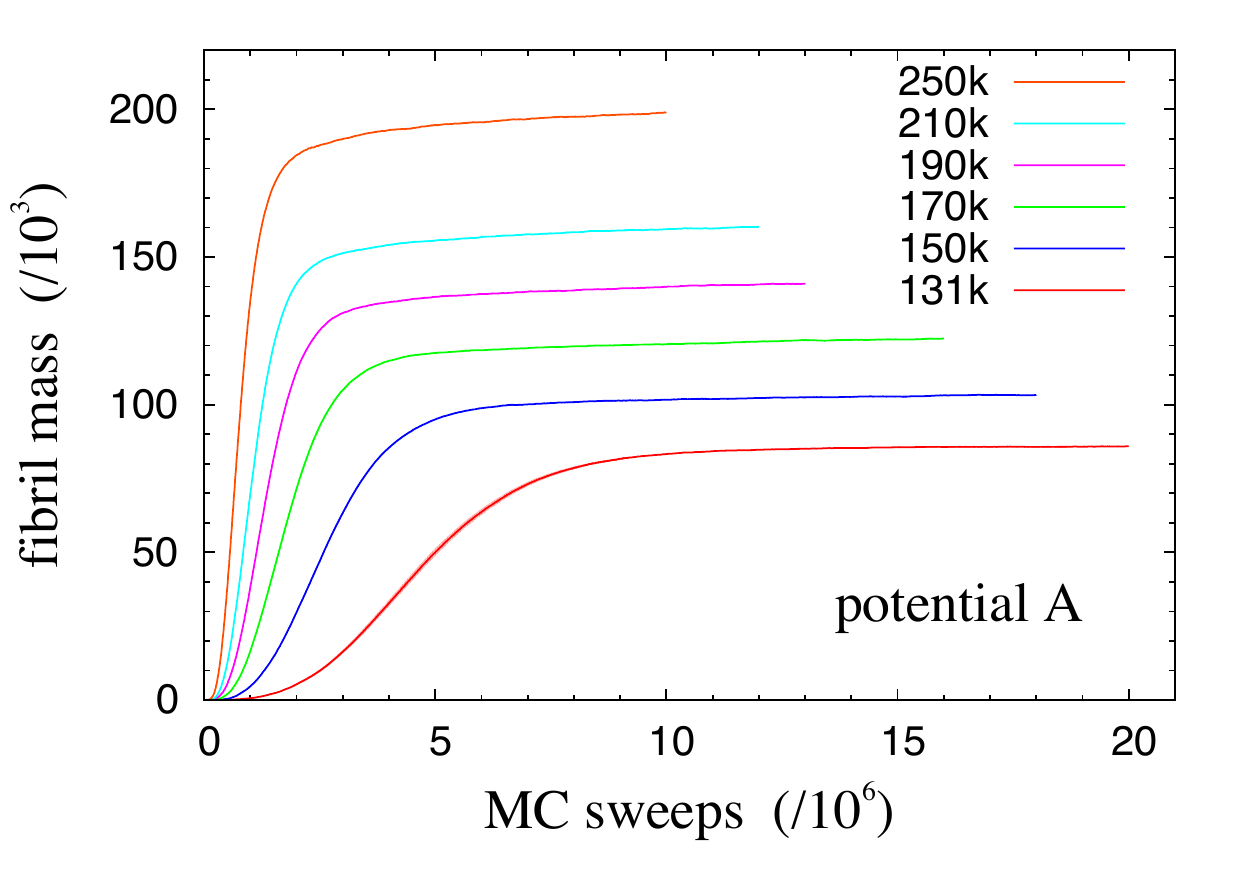}
  \includegraphics[width=8cm]{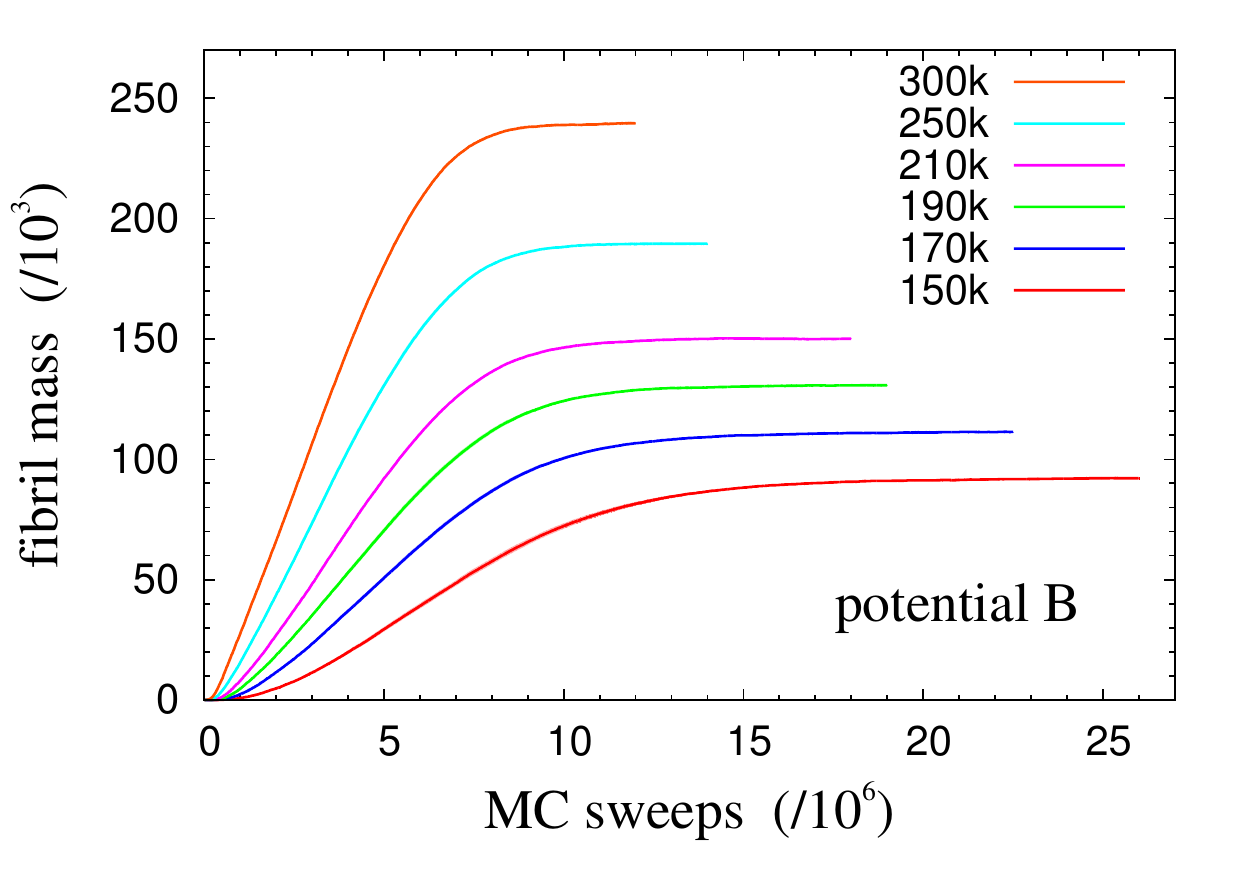}
  \end{center}
  \caption{
  MC evolution of the total fibril mass in relaxation simulations with 
  the potentials A and B (Table~\ref{tab:b}) for $T=0.6563$, $L=512$ and 
  different $N$ between $2^{17}=131,072$ and 300,000. 
  The data represent averages over eight independent runs, started
  from random initial states.
  The statistical uncertainties, indicated by shaded bands, are barely visible.
  \label{fig:fibkin}}
\end{figure}

\begin{figure}[t]
\centering
  \includegraphics[width=8cm]{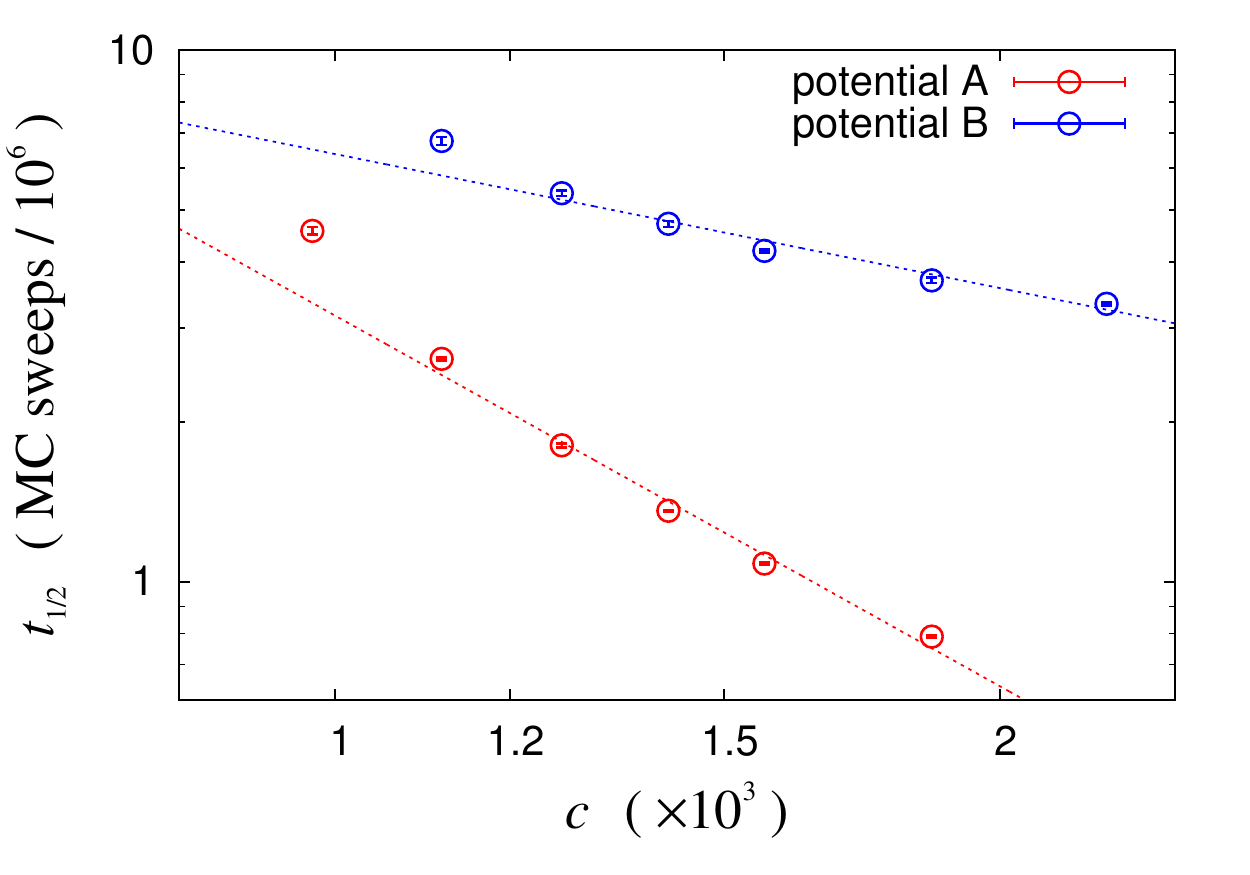}
  \caption{
  Concentration-dependence of the half-time for fibril formation, $\thalf$, 
  as extracted from the simulations shown in Fig.~\ref{fig:fibkin}. 
  The lines are power-law fits to the data, $\thalf\propto c^\gamma$. 
  The fitted exponents are $\gamma^\text{(A)}=-2.3\pm0.2$ and 
  $\gamma^\text{(B)}=-0.8\pm0.1$ for potentials A and B, respectively.
  \label{fig:thalf}}
\end{figure}

Fig.~\ref{fig:fibkin} 
shows the MC evolution of the total fibril mass 
in systems A and B for different concentrations. 
The curves are sigmoidal in shape, especially 
at low $c$. As expected, as $c$ is increased, 
aggregation gets faster and the saturation 
level gets higher. The statistical errors are small 
because our systems are large. A simple measure 
of the overall rate of fibril formation is the time, 
$\thalf$, at which half the saturation level is reached. 
In amyloid formation, the scaling of $\thalf$ with $c$   
has often, but not always,\cite{Meisl:14} 
been found to be well described by a power law, 
$\thalf\sim c^\gamma$, where the exponent $\gamma\le-0.5$ 
depends on both the protein and the conditions under which
the fibrils grow.\cite{Knowles:09} Data for $\thalf$ from 
our relaxation simulations do not show a perfect power-law
behavior, as can be seen from a log-log plot (Fig.~\ref{fig:thalf}).
Nevertheless, to get a measure of the overall strength of the 
$c$-dependence, power-law fits were performed. 
For system B, with quasi-2D growth, the fitted 
exponent, $\gamma^\text{(B)}=-0.8\pm0.1$, indicates a 
$c$-dependence comparable in strength to that of  
typical experimental data.\cite{Knowles:09} 
For system A, with 2D growth, the $c$-dependence is 
slightly stronger, with a fitted exponent 
of $\gamma^\text{(A)}=-2.3\pm0.2$.
            
\section{\label{sec:discussion}Summary}

Amyloid formation involves a wide range of spatial and temporal scales.
In this article, we have used a minimal lattice-based model to investigate the 
overall thermodynamics of amyloid formation in finite systems 
under $NVT$ conditions. With 2D or quasi-2D aggregate growth, the model 
exhibits a sharp transition, from a supersaturated solution state to a distinct 
state where small and large species exist in equilibrium. At the threshold 
concentration, $\cf(T)$, these states coexist, thus giving rise to a bimodal 
energy distribution. At concentrations not too much higher 
than $\cf(T)$, there exists, therefore, a local free-energy minimum 
corresponding to a metastable solution state, in which the system can get 
trapped, thereby causing fibril formation to occur after a lag period.  
At and above $\cf(T)$, while both small and large 
aggregates are present, intermediate-size ones are 
suppressed. With 1D growth, this suppression is not 
observed, and the energy distribution is unimodal. 
Intuitively, the first-order-like transition seen with 2D or quasi-2D 
growth stems from a competition between bulk and surface energies. 
With 1D growth, this mechanism is missing, because the surface energy 
is associated with the fibril endpoints, whose size does not grow with 
fibril mass. Previous work has studied the dependence of the 
solubility of fibrils on their width, using different models.~\cite{Zhang:09,Auer:10} 
One study compared one-, two- and three-layered aggregates, 
and showed that the stability region in the  
$T$,$c$ plane grows with increasing fibril width.~\cite{Auer:10} 
This behavior suggests that fibril formation in a finite system 
may set in at a concentration roughly corresponding to the solubility 
of the widest aggregates that occur for this system size. 
Upon increasing $N$ (at fixed $c$ and $T$), one would then 
expect a growth in both latent heat and threshold concentration,
as is indeed observed in our simulations.    
 
The first-order-like onset of fibril formation that we observe with 2D or quasi-2D growth 
shows similarities with the droplet evaporation/condensation transition at liquid-vapor 
coexistence, which has been extensively investigated.\cite{Binder:80,Furukawa:82,Biskup:02,Biskup:03,
Neuhaus:03,Biskup:04,MacDowell:04,Nussbaumer:06,Nussbaumer:10,Bauer:10,Nogawa:11}    
Indeed, at this transition, mid-size droplets are suppressed and the energy distribution 
is bimodal. Furthermore, the specific latent heat, which we find to decrease with system 
size (Fig.~\ref{fig:fss}b),  
is known to vanish at the droplet transition in the limit of infinite system size. 

Our equilibrium findings may be used to 
rationalize, in part, properties observed in our relaxation simulations. For the
systems studied here, with $>$$10^5$ peptides, the MC evolution 
of the total fibril mass turns out to be highly reproducible from run to run. The
trajectories are, at not too high concentrations, sigmoidal, 
with an initial lag phase (Fig.~\ref{fig:fibkin}). Due to slow dynamics of large aggregates,
the apparent steady-state levels at the end of the runs need not correspond to 
equilibrium states. However, as at equilibrium, intermediate-size 
aggregates are suppressed in the final states (Fig.~\ref{fig:massdist}),
which is in line with experimental findings.~\cite{Walsh:97}.
With the droplet interpretation, this statistical suppression
occurs because intermediate-size aggregates correspond to a free-energy maximum, 
at which the bulk free energy and surface energy terms balance each 
other. The precise shape of the aggregate-size distribution is influenced
by factors that are unlikely to be captured by our simple model, such 
as the existence of specific oligomeric states with enhanced stability. 
One type of aggregate that does not occur in our simulations, due to  
the model geometry, is closed $\beta$-barrels, which have a 
potentially high stability for their size.~\cite{Irback:08}

\begin{acknowledgments}
This work benefitted greatly from the expertise and helpfulness
of the late Thomas Neuhaus.
We thank Sigur\dh ur \AE. J\'onsson and Stefan Wallin for useful
discussions. This work was in part supported by the Swedish 
Research Council (Grant no. 621-2014-4522). 
The simulations were performed on resources provided by the Swedish
National Infrastructure for Computing (SNIC) at LUNARC, Lund
University.
\end{acknowledgments}


\begin{thebibliography}{10}%
\makeatletter
\providecommand \@ifxundefined [1]{%
 \ifx #1\undefined \expandafter \@firstoftwo
 \else \expandafter \@secondoftwo
\fi
}%
\providecommand \@ifnum [1]{%
 \ifnum #1\expandafter \@firstoftwo
 \else \expandafter \@secondoftwo
\fi
}%
\providecommand \enquote [1]{``#1''}%
\providecommand \bibnamefont  [1]{#1}%
\providecommand \bibfnamefont [1]{#1}%
\providecommand \citenamefont [1]{#1}%
\providecommand\href[0]{\@sanitize\@href}%
\providecommand\@href[1]{\endgroup\@@startlink{#1}\endgroup\@@href}%
\providecommand\@@href[1]{#1\@@endlink}%
\providecommand \@sanitize [0]{\begingroup\catcode`\&12\catcode`\#12\relax}%
\@ifxundefined \pdfoutput {\@firstoftwo}{%
 \@ifnum{\z@=\pdfoutput}{\@firstoftwo}{\@secondoftwo}%
}{%
 \providecommand\@@startlink[1]{\leavevmode}%
 \providecommand\@@endlink[0]{}%
}{%
 \providecommand\@@startlink[1]{%
  \leavevmode
  \pdfstartlink
   attr{/Border[0 0 1 ]/H/I/C[0 1 1]}%
   user{/Subtype/Link/A<</Type/Action/S/URI/URI(#1)>>}%
  \relax
 }%
 \providecommand\@@endlink[0]{\pdfendlink}%
}%
\providecommand \url  [0]{\begingroup\@sanitize \@url }%
\providecommand \@url [1]{\endgroup\@href {#1}{\urlprefix}}%
\providecommand \urlprefix [0]{URL }%
\providecommand \Eprint[0]{\href }%
\@ifxundefined \urlstyle {%
  \providecommand \doi [1]{doi:\discretionary{}{}{}#1}%
}{%
  \providecommand \doi [0]{doi:\discretionary{}{}{}\begingroup
  \urlstyle{rm}\Url }%
}%
\providecommand \doibase [0]{http://dx.doi.org/}%
\providecommand \Doi[1]{\href{\doibase#1}}%
\providecommand \selectlanguage [0]{\@gobble}%
\providecommand \bibinfo [0]{\@secondoftwo}%
\providecommand \bibfield [0]{\@secondoftwo}%
\providecommand \translation [1]{[#1]}%
\providecommand \BibitemOpen[0]{}%
\providecommand \bibitemStop [0]{}%
\providecommand \bibitemNoStop [0]{.\EOS\space}%
\providecommand \EOS [0]{\spacefactor3000\relax}%
\providecommand \BibitemShut [1]{\csname bibitem#1\endcsname}%
\bibitem{Chiti:06}%
  \BibitemOpen
  \bibfield{author}{%
  \bibinfo {author} {\bibfnamefont{F.}~\bibnamefont{Chiti}}\ and\ \bibinfo
  {author} {\bibfnamefont{C.~M.}\ \bibnamefont{Dobson}},\ }%
  \bibfield{journal}{%
  \bibinfo {journal} {Annu.\ Rev.\ Biochem.}\ }%
  \textbf{\bibinfo {volume} {75}},\ \bibinfo {pages} {333} (\bibinfo {year}
  {2006})\BibitemShut{NoStop}%
\bibitem{Knowles:11}%
  \BibitemOpen
  \bibfield{author}{%
  \bibinfo {author} {\bibfnamefont{T.~P.~J.}\ \bibnamefont{Knowles}}\ and\
  \bibinfo {author} {\bibfnamefont{M.~J.}\ \bibnamefont{Buehler}},\ }%
  \bibfield{journal}{%
  \bibinfo {journal} {Nat.\ Nanotechnol.}\ }%
  \textbf{\bibinfo {volume} {6}},\ \bibinfo {pages} {469} (\bibinfo {year}
  {2011})\BibitemShut{NoStop}%
\bibitem{Hard:14}%
  \BibitemOpen
  \bibfield{author}{%
  \bibinfo {author} {\bibfnamefont{T.}~\bibnamefont{H{\"a}rd}},\ }%
  \bibfield{journal}{%
  \bibinfo {journal} {J.\ Phys.\ Chem.\ Lett.}\ }%
  \textbf{\bibinfo {volume} {5}},\ \bibinfo {pages} {607} (\bibinfo {year}
  {2014})\BibitemShut{NoStop}%
\bibitem{Hellstrand:10}%
  \BibitemOpen
  \bibfield{author}{%
  \bibinfo {author} {\bibfnamefont{E.}~\bibnamefont{Hellstrand}}, \bibinfo
  {author} {\bibfnamefont{B.}~\bibnamefont{Boland}}, \bibinfo {author}
  {\bibfnamefont{D.~M.}\ \bibnamefont{Walsh}},\ and\ \bibinfo {author}
  {\bibfnamefont{S.}~\bibnamefont{Linse}},\ }%
  \bibfield{journal}{%
  \bibinfo {journal} {ACS\ Chem.\ Neurosci.}\ }%
  \textbf{\bibinfo {volume} {1}},\ \bibinfo {pages} {13} (\bibinfo {year}
  {2010})\BibitemShut{NoStop}%
\bibitem{Knowles:09}%
  \BibitemOpen
  \bibfield{author}{%
  \bibinfo {author} {\bibfnamefont{T.~P.~J.}\ \bibnamefont{Knowles}}, \bibinfo
  {author} {\bibfnamefont{C.~A.}\ \bibnamefont{Waudby}}, \bibinfo {author}
  {\bibfnamefont{G.~L.}\ \bibnamefont{Devlin}}, \bibinfo {author}
  {\bibfnamefont{S.~I.~A.}\ \bibnamefont{Cohen}}, \bibinfo {author}
  {\bibfnamefont{A.}~\bibnamefont{Aguzzi}}, \bibinfo {author}
  {\bibfnamefont{M.}~\bibnamefont{Vendruscolo}}, \bibinfo {author}
  {\bibfnamefont{E.~M.}\ \bibnamefont{Terentjev}}, \bibinfo {author}
  {\bibfnamefont{M.~E.}\ \bibnamefont{Welland}},\ and\ \bibinfo {author}
  {\bibfnamefont{C.~M.}\ \bibnamefont{Dobson}},\ }%
  \bibfield{journal}{%
  \bibinfo {journal} {Science}\ }%
  \textbf{\bibinfo {volume} {326}},\ \bibinfo {pages} {1533} (\bibinfo {year}
  {2009})\BibitemShut{NoStop}%
\bibitem{Ferrone:85}%
  \BibitemOpen
  \bibfield{author}{%
  \bibinfo {author} {\bibfnamefont{F.~A.}\ \bibnamefont{Ferrone}}, \bibinfo
  {author} {\bibfnamefont{J.}~\bibnamefont{Hofrichter}},\ and\ \bibinfo
  {author} {\bibfnamefont{W.}~\bibnamefont{Eaton}},\ }%
  \bibfield{journal}{%
  \bibinfo {journal} {J.\ Mol.\ Biol.}\ }%
  \textbf{\bibinfo {volume} {183}},\ \bibinfo {pages} {611} (\bibinfo {year}
  {1985})\BibitemShut{NoStop}%
\bibitem{Flyvbjerg:96}%
  \BibitemOpen
  \bibfield{author}{%
  \bibinfo {author} {\bibfnamefont{H.}~\bibnamefont{Flyvbjerg}}, \bibinfo
  {author} {\bibfnamefont{E.}~\bibnamefont{Jobs}},\ and\ \bibinfo {author}
  {\bibfnamefont{S.}~\bibnamefont{Leibler}},\ }%
  \bibfield{journal}{%
  \bibinfo {journal} {Proc.\ Natl.\ Acad.\ Sci.\ USA}\ }%
  \textbf{\bibinfo {volume} {93}},\ \bibinfo {pages} {5975} (\bibinfo {year}
  {1996})\BibitemShut{NoStop}%
\bibitem{Straub:10}%
  \BibitemOpen
  \bibfield{author}{%
  \bibinfo {author} {\bibfnamefont{J.~E.}\ \bibnamefont{Straub}}\ and\ \bibinfo
  {author} {\bibfnamefont{D.}~\bibnamefont{Thirumalai}},\ }%
  \bibfield{journal}{%
  \bibinfo {journal} {Curr.\ Opin.\ Struct.\ Biol.}\ }%
  \textbf{\bibinfo {volume} {20}},\ \bibinfo {pages} {187} (\bibinfo {year}
  {2010})\BibitemShut{NoStop}%
\bibitem{Auer:08}%
  \BibitemOpen
  \bibfield{author}{%
  \bibinfo {author} {\bibfnamefont{S.}~\bibnamefont{Auer}}, \bibinfo {author}
  {\bibfnamefont{C.~M.}\ \bibnamefont{Dobson}}, \bibinfo {author}
  {\bibfnamefont{M.}~\bibnamefont{Vendruscolo}},\ and\ \bibinfo {author}
  {\bibfnamefont{A.}~\bibnamefont{Maritan}},\ }%
  \bibfield{journal}{%
  \bibinfo {journal} {Phys.\ Rev.\ Lett.}\ }%
  \textbf{\bibinfo {volume} {101}},\ \bibinfo {pages} {258101} (\bibinfo {year}
  {2008})\BibitemShut{NoStop}%
\bibitem{Junghans:08}%
  \BibitemOpen
  \bibfield{author}{%
  \bibinfo {author} {\bibfnamefont{C.}~\bibnamefont{Junghans}}, \bibinfo
  {author} {\bibfnamefont{M.}~\bibnamefont{Bachmann}},\ and\ \bibinfo {author}
  {\bibfnamefont{W.}~\bibnamefont{Janke}},\ }%
  \bibfield{journal}{%
  \bibinfo {journal} {J.\ Chem.\ Phys.}\ }%
  \textbf{\bibinfo {volume} {128}},\ \bibinfo {pages} {085103} (\bibinfo {year}
  {2008})\BibitemShut{NoStop}%
\bibitem{Irback:08}%
  \BibitemOpen
  \bibfield{author}{%
  \bibinfo {author} {\bibfnamefont{A.}~\bibnamefont{Irb{\"a}ck}}\ and\ \bibinfo
  {author} {\bibfnamefont{S.}~\bibnamefont{Mitternacht}},\ }%
  \bibfield{journal}{%
  \bibinfo {journal} {Proteins}\ }%
  \textbf{\bibinfo {volume} {71}},\ \bibinfo {pages} {207} (\bibinfo {year}
  {2008})\BibitemShut{NoStop}%
\bibitem{Li:08}%
  \BibitemOpen
  \bibfield{author}{%
  \bibinfo {author} {\bibfnamefont{D.}~\bibnamefont{Li}}, \bibinfo {author}
  {\bibfnamefont{S.}~\bibnamefont{Mohanty}}, \bibinfo {author}
  {\bibfnamefont{A.}~\bibnamefont{Irb\"ack}},\ and\ \bibinfo {author}
  {\bibfnamefont{S.}~\bibnamefont{Huo}},\ }%
  \bibfield{journal}{%
  \bibinfo {journal} {PLoS\ Comput.\ Biol.}\ }%
  \textbf{\bibinfo {volume} {4}},\ \bibinfo {pages} {e1000238} (\bibinfo {year}
  {2008})\BibitemShut{NoStop}%
\bibitem{LiMS:08}%
  \BibitemOpen
  \bibfield{author}{%
  \bibinfo {author} {\bibfnamefont{M.~S.}\ \bibnamefont{Li}}, \bibinfo {author}
  {\bibfnamefont{D.~K.}\ \bibnamefont{Klimov}}, \bibinfo {author}
  {\bibfnamefont{J.~E.}\ \bibnamefont{Straub}},\ and\ \bibinfo {author}
  {\bibfnamefont{D.}~\bibnamefont{Thirumalai}},\ }%
  \bibfield{journal}{%
  \bibinfo {journal} {J.\ Chem.\ Phys.}\ }%
  \textbf{\bibinfo {volume} {129}},\ \bibinfo {pages} {175101} (\bibinfo {year}
  {2008})\BibitemShut{NoStop}%
\bibitem{Bellesia:09}%
  \BibitemOpen
  \bibfield{author}{%
  \bibinfo {author} {\bibfnamefont{G.}~\bibnamefont{Bellesia}}\ and\ \bibinfo
  {author} {\bibfnamefont{J.-E.}\ \bibnamefont{Shea}},\ }%
  \bibfield{journal}{%
  \bibinfo {journal} {J.\ Chem.\ Phys.}\ }%
  \textbf{\bibinfo {volume} {130}},\ \bibinfo {pages} {145103} (\bibinfo {year}
  {2009})\BibitemShut{NoStop}%
\bibitem{Lu:09}%
  \BibitemOpen
  \bibfield{author}{%
  \bibinfo {author} {\bibfnamefont{Y.}~\bibnamefont{Lu}}, \bibinfo {author}
  {\bibfnamefont{P.}~\bibnamefont{Derreumaux}}, \bibinfo {author}
  {\bibfnamefont{Z.}~\bibnamefont{Guo}}, \bibinfo {author}
  {\bibfnamefont{N.}~\bibnamefont{Mousseau}},\ and\ \bibinfo {author}
  {\bibfnamefont{G.}~\bibnamefont{Wei}},\ }%
  \bibfield{journal}{%
  \bibinfo {journal} {Proteins}\ }%
  \textbf{\bibinfo {volume} {75}},\ \bibinfo {pages} {954} (\bibinfo {year}
  {2009})\BibitemShut{NoStop}%
\bibitem{Wang:10}%
  \BibitemOpen
  \bibfield{author}{%
  \bibinfo {author} {\bibfnamefont{Y.}~\bibnamefont{Wang}}\ and\ \bibinfo
  {author} {\bibfnamefont{G.~A.}\ \bibnamefont{Voth}},\ }%
  \bibfield{journal}{%
  \bibinfo {journal} {J.\ Phys.\ Chem.\ B}\ }%
  \textbf{\bibinfo {volume} {114}},\ \bibinfo {pages} {8735} (\bibinfo {year}
  {2010})\BibitemShut{NoStop}%
\bibitem{Auer:10}%
  \BibitemOpen
  \bibfield{author}{%
  \bibinfo {author} {\bibfnamefont{S.}~\bibnamefont{Auer}}\ and\ \bibinfo
  {author} {\bibfnamefont{D.}~\bibnamefont{Kashchiev}},\ }%
  \bibfield{journal}{%
  \bibinfo {journal} {Phys.\ Rev.\ Lett.}\ }%
  \textbf{\bibinfo {volume} {104}},\ \bibinfo {pages} {168105} (\bibinfo {year}
  {2010})\BibitemShut{NoStop}%
\bibitem{Kashchiev:10}%
  \BibitemOpen
  \bibfield{author}{%
  \bibinfo {author} {\bibfnamefont{D.}~\bibnamefont{Kashchiev}}\ and\ \bibinfo
  {author} {\bibfnamefont{S.}~\bibnamefont{Auer}},\ }%
  \bibfield{journal}{%
  \bibinfo {journal} {J.\ Chem.\ Phys.}\ }%
  \textbf{\bibinfo {volume} {132}},\ \bibinfo {pages} {215101} (\bibinfo {year}
  {2010})\BibitemShut{NoStop}%
\bibitem{Friedman:10}%
  \BibitemOpen
  \bibfield{author}{%
  \bibinfo {author} {\bibfnamefont{R.}~\bibnamefont{Friedman}}, \bibinfo
  {author} {\bibfnamefont{R.}~\bibnamefont{Pellarin}},\ and\ \bibinfo {author}
  {\bibfnamefont{A.}~\bibnamefont{Caflisch}},\ }%
  \bibfield{journal}{%
  \bibinfo {journal} {J.\ Phys.\ Chem.\ Lett.}\ }%
  \textbf{\bibinfo {volume} {1}},\ \bibinfo {pages} {471} (\bibinfo {year}
  {2010})\BibitemShut{NoStop}%
\bibitem{Rojas:10}%
  \BibitemOpen
  \bibfield{author}{%
  \bibinfo {author} {\bibfnamefont{A.}~\bibnamefont{Rojas}}, \bibinfo {author}
  {\bibfnamefont{A.}~\bibnamefont{Liwo}}, \bibinfo {author}
  {\bibfnamefont{D.}~\bibnamefont{Browne}},\ and\ \bibinfo {author}
  {\bibfnamefont{H.~A.}\ \bibnamefont{Scheraga}},\ }%
  \bibfield{journal}{%
  \bibinfo {journal} {J.\ Mol.\ Biol.}\ }%
  \textbf{\bibinfo {volume} {404}},\ \bibinfo {pages} {537} (\bibinfo {year}
  {2010})\BibitemShut{NoStop}%
\bibitem{Urbanc:10}%
  \BibitemOpen
  \bibfield{author}{%
  \bibinfo {author} {\bibfnamefont{B.}~\bibnamefont{Urbanc}}, \bibinfo {author}
  {\bibfnamefont{M.}~\bibnamefont{Betnel}}, \bibinfo {author}
  {\bibfnamefont{L.}~\bibnamefont{Cruz}}, \bibinfo {author}
  {\bibfnamefont{G.}~\bibnamefont{Bitan}},\ and\ \bibinfo {author}
  {\bibfnamefont{D.~B.}\ \bibnamefont{Teplow}},\ }%
  \bibfield{journal}{%
  \bibinfo {journal} {J.\ Am.\ Chem.\ Soc.}\ }%
  \textbf{\bibinfo {volume} {132}},\ \bibinfo {pages} {4266} (\bibinfo {year}
  {2010})\BibitemShut{NoStop}%
\bibitem{Kim:10}%
  \BibitemOpen
  \bibfield{author}{%
  \bibinfo {author} {\bibfnamefont{S.}~\bibnamefont{Kim}}, \bibinfo {author}
  {\bibfnamefont{T.}~\bibnamefont{Takeda}},\ and\ \bibinfo {author}
  {\bibfnamefont{D.~K.}\ \bibnamefont{Klimov}},\ }%
  \bibfield{journal}{%
  \bibinfo {journal} {Biophys.\ J.}\ }%
  \textbf{\bibinfo {volume} {99}},\ \bibinfo {pages} {1949} (\bibinfo {year}
  {2010})\BibitemShut{NoStop}%
\bibitem{Cheon:11}%
  \BibitemOpen
  \bibfield{author}{%
  \bibinfo {author} {\bibfnamefont{M.}~\bibnamefont{Cheon}}, \bibinfo {author}
  {\bibfnamefont{I.}~\bibnamefont{Chang}},\ and\ \bibinfo {author}
  {\bibfnamefont{C.~K.}\ \bibnamefont{Hall}},\ }%
  \bibfield{journal}{%
  \bibinfo {journal} {Biophys.\ J.}\ }%
  \textbf{\bibinfo {volume} {101}},\ \bibinfo {pages} {2493} (\bibinfo {year}
  {2011})\BibitemShut{NoStop}%
\bibitem{Linse:11}%
  \BibitemOpen
  \bibfield{author}{%
  \bibinfo {author} {\bibfnamefont{B.}~\bibnamefont{Linse}}\ and\ \bibinfo
  {author} {\bibfnamefont{S.}~\bibnamefont{Linse}},\ }%
  \bibfield{journal}{%
  \bibinfo {journal} {Mol.\ BioSyst.}\ }%
  \textbf{\bibinfo {volume} {7}},\ \bibinfo {pages} {2296} (\bibinfo {year}
  {2011})\BibitemShut{NoStop}%
\bibitem{Baiesi:11}%
  \BibitemOpen
  \bibfield{author}{%
  \bibinfo {author} {\bibfnamefont{M.}~\bibnamefont{Baiesi}}, \bibinfo {author}
  {\bibfnamefont{F.}~\bibnamefont{Seno}},\ and\ \bibinfo {author}
  {\bibfnamefont{A.}~\bibnamefont{Trovato}},\ }%
  \bibfield{journal}{%
  \bibinfo {journal} {Proteins}\ }%
  \textbf{\bibinfo {volume} {79}},\ \bibinfo {pages} {3067} (\bibinfo {year}
  {2011})\BibitemShut{NoStop}%
\bibitem{Carmichael:12}%
  \BibitemOpen
  \bibfield{author}{%
  \bibinfo {author} {\bibfnamefont{S.~P.}\ \bibnamefont{Carmichael}}\ and\
  \bibinfo {author} {\bibfnamefont{M.~S.}\ \bibnamefont{Shell}},\ }%
  \bibfield{journal}{%
  \bibinfo {journal} {J.\ Phys.\ Chem.\ B}\ }%
  \textbf{\bibinfo {volume} {116}},\ \bibinfo {pages} {8383} (\bibinfo {year}
  {2012})\BibitemShut{NoStop}%
\bibitem{Bieler:12}%
  \BibitemOpen
  \bibfield{author}{%
  \bibinfo {author} {\bibfnamefont{N.~S.}\ \bibnamefont{Bieler}}, \bibinfo
  {author} {\bibfnamefont{T.~P.~J.}\ \bibnamefont{Knowles}}, \bibinfo {author}
  {\bibfnamefont{D.}~\bibnamefont{Frenkel}},\ and\ \bibinfo {author}
  {\bibfnamefont{R.}~\bibnamefont{V{\'a}cha}},\ }%
  \bibfield{journal}{%
  \bibinfo {journal} {PLoS\ Comput.\ Biol.}\ }%
  \textbf{\bibinfo {volume} {8}},\ \bibinfo {pages} {e1002692} (\bibinfo {year}
  {2012})\BibitemShut{NoStop}%
\bibitem{Smaoui:13}%
  \BibitemOpen
  \bibfield{author}{%
  \bibinfo {author} {\bibfnamefont{M.~R.}\ \bibnamefont{Smaoui}}, \bibinfo
  {author} {\bibfnamefont{F.}~\bibnamefont{Poitevin}}, \bibinfo {author}
  {\bibfnamefont{M.}~\bibnamefont{Delarue}}, \bibinfo {author}
  {\bibfnamefont{P.}~\bibnamefont{Koehl}}, \bibinfo {author}
  {\bibfnamefont{H.}~\bibnamefont{Orland}},\ and\ \bibinfo {author}
  {\bibfnamefont{J.}~\bibnamefont{Waldisp{\"u}hl}},\ }%
  \bibfield{journal}{%
  \bibinfo {journal} {Biophys.\ J.}\ }%
  \textbf{\bibinfo {volume} {104}},\ \bibinfo {pages} {683} (\bibinfo {year}
  {2013})\BibitemShut{NoStop}%
\bibitem{Ni:13}%
  \BibitemOpen
  \bibfield{author}{%
  \bibinfo {author} {\bibfnamefont{R.}~\bibnamefont{Ni}}, \bibinfo {author}
  {\bibfnamefont{S.}~\bibnamefont{Abeln}}, \bibinfo {author}
  {\bibfnamefont{M.}~\bibnamefont{Schor}}, \bibinfo {author}
  {\bibfnamefont{M.~A.}\ \bibnamefont{Cohen~Stuart}},\ and\ \bibinfo {author}
  {\bibfnamefont{P.~G.}\ \bibnamefont{Bolhuis}},\ }%
  \bibfield{journal}{%
  \bibinfo {journal} {Phys.\ Rev.\ Lett.}\ }%
  \textbf{\bibinfo {volume} {111}},\ \bibinfo {pages} {058101} (\bibinfo {year}
  {2013})\BibitemShut{NoStop}%
\bibitem{Zheng:13}%
  \BibitemOpen
  \bibfield{author}{%
  \bibinfo {author} {\bibfnamefont{W.}~\bibnamefont{Zheng}}, \bibinfo {author}
  {\bibfnamefont{N.~P.}\ \bibnamefont{Schafer}},\ and\ \bibinfo {author}
  {\bibfnamefont{P.~G.}\ \bibnamefont{Wolynes}},\ }%
  \bibfield{journal}{%
  \bibinfo {journal} {Proc.\ Natl.\ Acad.\ Sci.\ USA}\ }%
  \textbf{\bibinfo {volume} {110}},\ \bibinfo {pages} {20515} (\bibinfo {year}
  {2013})\BibitemShut{NoStop}%
\bibitem{DiMichele:13}%
  \BibitemOpen
  \bibfield{author}{%
  \bibinfo {author} {\bibfnamefont{L.}~\bibnamefont{Di~Michele}}, \bibinfo
  {author} {\bibfnamefont{E.}~\bibnamefont{Eiser}},\ and\ \bibinfo {author}
  {\bibfnamefont{V.}~\bibnamefont{Foder{\`a}}},\ }%
  \bibfield{journal}{%
  \bibinfo {journal} {J.\ Phys.\ Chem.\ Lett.}\ }%
  \textbf{\bibinfo {volume} {4}},\ \bibinfo {pages} {3158} (\bibinfo {year}
  {2013})\BibitemShut{NoStop}%
\bibitem{Abeln:14}%
  \BibitemOpen
  \bibfield{author}{%
  \bibinfo {author} {\bibfnamefont{S.}~\bibnamefont{Abeln}}, \bibinfo {author}
  {\bibfnamefont{M.}~\bibnamefont{Vendruscolo}}, \bibinfo {author}
  {\bibfnamefont{C.~M.}\ \bibnamefont{Dobson}},\ and\ \bibinfo {author}
  {\bibfnamefont{D.}~\bibnamefont{Frenkel}},\ }%
  \bibfield{journal}{%
  \bibinfo {journal} {PLoS\ One}\ }%
  \textbf{\bibinfo {volume} {9}},\ \bibinfo {pages} {e85185} (\bibinfo {year}
  {2014})\BibitemShut{NoStop}%
\bibitem{MorrissAndrews:14}%
  \BibitemOpen
  \bibfield{author}{%
  \bibinfo {author} {\bibfnamefont{A.}~\bibnamefont{Morriss-Andrews}}\ and\
  \bibinfo {author} {\bibfnamefont{J.-E.}\ \bibnamefont{Shea}},\ }%
  \bibfield{journal}{%
  \bibinfo {journal} {J.\ Phys.\ Chem.\ Lett.}\ }%
  \textbf{\bibinfo {volume} {5}},\ \bibinfo {pages} {1899} (\bibinfo {year}
  {2014})\BibitemShut{NoStop}%
\bibitem{Saric:14}%
  \BibitemOpen
  \bibfield{author}{%
  \bibinfo {author} {\bibfnamefont{A.}~\bibnamefont{{\v{S}}ari{\'c}}}, \bibinfo
  {author} {\bibfnamefont{Y.~C.}\ \bibnamefont{Chebaro}}, \bibinfo {author}
  {\bibfnamefont{T.~P.~J.}\ \bibnamefont{Knowles}},\ and\ \bibinfo {author}
  {\bibfnamefont{D.}~\bibnamefont{Frenkel}},\ }%
  \bibfield{journal}{%
  \bibinfo {journal} {Proc.\ Natl.\ Acad.\ Sci.\ USA}\ }%
  \textbf{\bibinfo {volume} {111}},\ \bibinfo {pages} {17869} (\bibinfo {year}
  {2014})\BibitemShut{NoStop}%
\bibitem{Assenza:14}%
  \BibitemOpen
  \bibfield{author}{%
  \bibinfo {author} {\bibfnamefont{S.}~\bibnamefont{Assenza}}, \bibinfo
  {author} {\bibfnamefont{J.}~\bibnamefont{Adamcik}}, \bibinfo {author}
  {\bibfnamefont{R.}~\bibnamefont{Mezzenga}},\ and\ \bibinfo {author}
  {\bibfnamefont{P.}~\bibnamefont{De~Los~Rios}},\ }%
  \bibfield{journal}{%
  \bibinfo {journal} {Phys.\ Rev.\ Lett.}\ }%
  \textbf{\bibinfo {volume} {113}},\ \bibinfo {pages} {268103} (\bibinfo {year}
  {2014})\BibitemShut{NoStop}%
\bibitem{Zhang:09}%
  \BibitemOpen
  \bibfield{author}{%
  \bibinfo {author} {\bibfnamefont{J.}~\bibnamefont{Zhang}}\ and\ \bibinfo
  {author} {\bibfnamefont{M.}~\bibnamefont{Muthukumar}},\ }%
  \bibfield{journal}{%
  \bibinfo {journal} {J.\ Chem.\ Phys.}\ }%
  \textbf{\bibinfo {volume} {130}},\ \bibinfo {pages} {035102} (\bibinfo {year}
  {2009})\BibitemShut{NoStop}%
\bibitem{Auer:11}%
  \BibitemOpen
  \bibfield{author}{%
  \bibinfo {author} {\bibfnamefont{S.}~\bibnamefont{Auer}},\ }%
  \bibfield{journal}{%
  \bibinfo {journal} {J.\ Chem.\ Phys.}\ }%
  \textbf{\bibinfo {volume} {135}},\ \bibinfo {pages} {175103} (\bibinfo {year}
  {2011})\BibitemShut{NoStop}%
\bibitem{Schmit:11}%
  \BibitemOpen
  \bibfield{author}{%
  \bibinfo {author} {\bibfnamefont{J.~D.}\ \bibnamefont{Schmit}}, \bibinfo
  {author} {\bibfnamefont{K.}~\bibnamefont{Ghosh}},\ and\ \bibinfo {author}
  {\bibfnamefont{K.}~\bibnamefont{Dill}},\ }%
  \bibfield{journal}{%
  \bibinfo {journal} {Biophys.\ J.}\ }%
  \textbf{\bibinfo {volume} {100}},\ \bibinfo {pages} {450} (\bibinfo {year}
  {2011})\BibitemShut{NoStop}%
\bibitem{Auer:14}%
  \BibitemOpen
  \bibfield{author}{%
  \bibinfo {author} {\bibfnamefont{S.}~\bibnamefont{Auer}},\ }%
  \bibfield{journal}{%
  \bibinfo {journal} {J.\ Phys.\ Chem.\ B}\ }%
  \textbf{\bibinfo {volume} {118}},\ \bibinfo {pages} {5289} (\bibinfo {year}
  {2014})\BibitemShut{NoStop}%
\bibitem{Swendsen:87}%
  \BibitemOpen
  \bibfield{author}{%
  \bibinfo {author} {\bibfnamefont{R.~H.}\ \bibnamefont{Swendsen}}\ and\
  \bibinfo {author} {\bibfnamefont{J.-S.}\ \bibnamefont{Wang}},\ }%
  \bibfield{journal}{%
  \bibinfo {journal} {Phys.\ Rev.\ Lett.}\ }%
  \textbf{\bibinfo {volume} {58}},\ \bibinfo {pages} {86} (\bibinfo {year}
  {1987})\BibitemShut{NoStop}%
\bibitem{Berg:91}%
  \BibitemOpen
  \bibfield{author}{%
  \bibinfo {author} {\bibfnamefont{B.~A.}\ \bibnamefont{Berg}}\ and\ \bibinfo
  {author} {\bibfnamefont{T.}~\bibnamefont{Neuhaus}},\ }%
  \bibfield{journal}{%
  \bibinfo {journal} {Phys.\ Lett.\ B}\ }%
  \textbf{\bibinfo {volume} {267}},\ \bibinfo {pages} {249} (\bibinfo {year}
  {1991})\BibitemShut{NoStop}%
\bibitem{Hansmann:93}%
  \BibitemOpen
  \bibfield{author}{%
  \bibinfo {author} {\bibfnamefont{U.~H.~E.}\ \bibnamefont{Hansmann}}\ and\
  \bibinfo {author} {\bibfnamefont{Y.}~\bibnamefont{Okamoto}},\ }%
  \bibfield{journal}{%
  \bibinfo {journal} {J.\ Comput.\ Chem.}\ }%
  \textbf{\bibinfo {volume} {14}},\ \bibinfo {pages} {1333} (\bibinfo {year}
  {1993})\BibitemShut{NoStop}%
\bibitem{Wang:01a}%
  \BibitemOpen
  \bibfield{author}{%
  \bibinfo {author} {\bibfnamefont{F.}~\bibnamefont{Wang}}\ and\ \bibinfo
  {author} {\bibfnamefont{D.~P.}\ \bibnamefont{Landau}},\ }%
  \bibfield{journal}{%
  \bibinfo {journal} {Phys.\ Rev.\ Lett.}\ }%
  \textbf{\bibinfo {volume} {86}},\ \bibinfo {pages} {2050} (\bibinfo {year}
  {2001})\BibitemShut{NoStop}%
\bibitem{Irback:13}%
  \BibitemOpen
  \bibfield{author}{%
  \bibinfo {author} {\bibfnamefont{A.}~\bibnamefont{Irb{\"a}ck}}, \bibinfo
  {author} {\bibfnamefont{S.~{\AE}.}\ \bibnamefont{J{\'o}nsson}}, \bibinfo
  {author} {\bibfnamefont{N.}~\bibnamefont{Linnemann}}, \bibinfo {author}
  {\bibfnamefont{B.}~\bibnamefont{Linse}},\ and\ \bibinfo {author}
  {\bibfnamefont{S.}~\bibnamefont{Wallin}},\ }%
  \bibfield{journal}{%
  \bibinfo {journal} {Phys.\ Rev.\ Lett.}\ }%
  \textbf{\bibinfo {volume} {110}},\ \bibinfo {pages} {058101} (\bibinfo {year}
  {2013})\BibitemShut{NoStop}%
\bibitem{Sawaya:07}%
  \BibitemOpen
  \bibfield{author}{%
  \bibinfo {author} {\bibfnamefont{M.~R.}\ \bibnamefont{Sawaya}}, \bibinfo
  {author} {\bibfnamefont{S.}~\bibnamefont{Sambashivan}}, \bibinfo {author}
  {\bibfnamefont{R.}~\bibnamefont{Nelson}}, \bibinfo {author}
  {\bibfnamefont{M.~I.}\ \bibnamefont{Ivanova}}, \bibinfo {author}
  {\bibfnamefont{S.~A.}\ \bibnamefont{Sievers}}, \bibinfo {author}
  {\bibfnamefont{M.~I.}\ \bibnamefont{Apostol}}, \bibinfo {author}
  {\bibfnamefont{M.~J.}\ \bibnamefont{Thompson}}, \bibinfo {author}
  {\bibfnamefont{M.}~\bibnamefont{Balbirnie}}, \bibinfo {author}
  {\bibfnamefont{J.~J.~W.}\ \bibnamefont{Wiltzius}}, \bibinfo {author}
  {\bibfnamefont{H.~T.}\ \bibnamefont{McFarlane}}, \bibinfo {author}
  {\bibfnamefont{A.~{\O}.}\ \bibnamefont{Madsen}}, \bibinfo {author}
  {\bibfnamefont{C.}~\bibnamefont{Riekel}},\ and\ \bibinfo {author}
  {\bibfnamefont{D.}~\bibnamefont{Eisenberg}},\ }%
  \bibfield{journal}{%
  \bibinfo {journal} {Nature}\ }%
  \textbf{\bibinfo {volume} {447}},\ \bibinfo {pages} {453} (\bibinfo {year}
  {2007})\BibitemShut{NoStop}%
\bibitem{Fitzpatrick:13}%
  \BibitemOpen
  \bibfield{author}{%
  \bibinfo {author} {\bibfnamefont{A.~W.~P.}\ \bibnamefont{Fitzpatrick}},
  \bibinfo {author} {\bibfnamefont{G.~T.}\ \bibnamefont{Debelouchina}},
  \bibinfo {author} {\bibfnamefont{M.~J.}\ \bibnamefont{Bayro}}, \bibinfo
  {author} {\bibfnamefont{D.~K.}\ \bibnamefont{Clare}}, \bibinfo {author}
  {\bibfnamefont{M.~A.}\ \bibnamefont{Caporini}}, \bibinfo {author}
  {\bibfnamefont{V.~S.}\ \bibnamefont{Bajaj}}, \bibinfo {author}
  {\bibfnamefont{C.~P.}\ \bibnamefont{Jaroniec}}, \bibinfo {author}
  {\bibfnamefont{L.}~\bibnamefont{Wang}}, \bibinfo {author}
  {\bibfnamefont{V.}~\bibnamefont{Ladizhansky}}, \bibinfo {author}
  {\bibfnamefont{S.~A.}\ \bibnamefont{M{\"u}ller}}, \bibinfo {author}
  {\bibfnamefont{C.~E.}\ \bibnamefont{MacPhee}}, \bibinfo {author}
  {\bibfnamefont{C.~A.}\ \bibnamefont{Waudby}}, \bibinfo {author}
  {\bibfnamefont{H.~R.}\ \bibnamefont{Mott}}, \bibinfo {author}
  {\bibfnamefont{A.}~\bibnamefont{De~Simone}}, \bibinfo {author}
  {\bibfnamefont{T.~P.~J.}\ \bibnamefont{Knowles}}, \bibinfo {author}
  {\bibfnamefont{H.~R.}\ \bibnamefont{Saibil}}, \bibinfo {author}
  {\bibfnamefont{M.}~\bibnamefont{Vendruscolo}}, \bibinfo {author}
  {\bibfnamefont{E.~V.}\ \bibnamefont{Orlova}}, \bibinfo {author}
  {\bibfnamefont{R.~G.}\ \bibnamefont{Griffin}},\ and\ \bibinfo {author}
  {\bibfnamefont{C.~M.}\ \bibnamefont{Dobson}},\ }%
  \bibfield{journal}{%
  \bibinfo {journal} {Proc.\ Natl.\ Acad.\ Sci.\ USA}\ }%
  \textbf{\bibinfo {volume} {110}},\ \bibinfo {pages} {5468} (\bibinfo {year}
  {2013})\BibitemShut{NoStop}%
\bibitem{Jonsson:11}%
  \BibitemOpen
  \bibfield{author}{%
  \bibinfo {author} {\bibfnamefont{S.~{\AE}.}\ \bibnamefont{J{\'o}nsson}},
  \bibinfo {author} {\bibfnamefont{S.}~\bibnamefont{Mohanty}},\ and\ \bibinfo
  {author} {\bibfnamefont{A.}~\bibnamefont{Irb{\"a}ck}},\ }%
  \bibfield{journal}{%
  \bibinfo {journal} {J.\ Chem.\ Phys.}\ }%
  \textbf{\bibinfo {volume} {135}},\ \bibinfo {pages} {125102} (\bibinfo {year}
  {2011})\BibitemShut{NoStop}%
\bibitem{Engkvist:96}%
  \BibitemOpen
  \bibfield{author}{%
  \bibinfo {author} {\bibfnamefont{O.}~\bibnamefont{Engkvist}}\ and\ \bibinfo
  {author} {\bibfnamefont{G.}~\bibnamefont{Karlstr{\"o}m}},\ }%
  \bibfield{journal}{%
  \bibinfo {journal} {Chem.\ Phys.}\ }%
  \textbf{\bibinfo {volume} {213}},\ \bibinfo {pages} {63} (\bibinfo {year}
  {1996})\BibitemShut{NoStop}%
\bibitem{Ferrenberg:89}%
  \BibitemOpen
  \bibfield{author}{%
  \bibinfo {author} {\bibfnamefont{A.~M.}\ \bibnamefont{Ferrenberg}}\ and\
  \bibinfo {author} {\bibfnamefont{R.~H.}\ \bibnamefont{Swendsen}},\ }%
  \bibfield{journal}{%
  \bibinfo {journal} {Phys.\ Rev.\ Lett.}\ }%
  \textbf{\bibinfo {volume} {63}},\ \bibinfo {pages} {1195} (\bibinfo {year}
  {1989})\BibitemShut{NoStop}%
\bibitem{Tian:14}%
  \BibitemOpen
  \bibfield{author}{%
  \bibinfo {author} {\bibfnamefont{P.}~\bibnamefont{Tian}}, \bibinfo {author}
  {\bibfnamefont{S.~{\AE}.}\ \bibnamefont{J{\'o}nsson}}, \bibinfo {author}
  {\bibfnamefont{J.}~\bibnamefont{Ferkinghoff-Borg}}, \bibinfo {author}
  {\bibfnamefont{S.~V.}\ \bibnamefont{Krivov}}, \bibinfo {author}
  {\bibfnamefont{K.}~\bibnamefont{Lindorff-Larsen}}, \bibinfo {author}
  {\bibfnamefont{A.}~\bibnamefont{Irb{\"a}ck}},\ and\ \bibinfo {author}
  {\bibfnamefont{W.}~\bibnamefont{Boomsma}},\ }%
  \bibfield{journal}{%
  \bibinfo {journal} {J.\ Chem.\ Theory\ Comput.}\ }%
  \textbf{\bibinfo {volume} {10}},\ \bibinfo {pages} {543} (\bibinfo {year}
  {2014})\BibitemShut{NoStop}%
\bibitem{Collins:04}%
  \BibitemOpen
  \bibfield{author}{%
  \bibinfo {author} {\bibfnamefont{S.~R.}\ \bibnamefont{Collins}}, \bibinfo
  {author} {\bibfnamefont{A.}~\bibnamefont{Douglass}}, \bibinfo {author}
  {\bibfnamefont{R.~D.}\ \bibnamefont{Vale}},\ and\ \bibinfo {author}
  {\bibfnamefont{J.~S.}\ \bibnamefont{Weissman}},\ }%
  \bibfield{journal}{%
  \bibinfo {journal} {PLoS\ Biol.}\ }%
  \textbf{\bibinfo {volume} {2}},\ \bibinfo {pages} {e321} (\bibinfo {year}
  {2004})\BibitemShut{NoStop}%
\bibitem{Oosawa:62}%
  \BibitemOpen
  \bibfield{author}{%
  \bibinfo {author} {\bibfnamefont{F.}~\bibnamefont{Oosawa}}\ and\ \bibinfo
  {author} {\bibfnamefont{M.}~\bibnamefont{Kasai}},\ }%
  \bibfield{journal}{%
  \bibinfo {journal} {J.\ Mol.\ Biol.}\ }%
  \textbf{\bibinfo {volume} {4}},\ \bibinfo {pages} {10} (\bibinfo {year}
  {1962})\BibitemShut{NoStop}%
\bibitem{Binder:80}%
  \BibitemOpen
  \bibfield{author}{%
  \bibinfo {author} {\bibfnamefont{K.}~\bibnamefont{Binder}}\ and\ \bibinfo
  {author} {\bibfnamefont{M.~H.}\ \bibnamefont{Kalos}},\ }%
  \bibfield{journal}{%
  \bibinfo {journal} {J.\ Stat.\ Phys.}\ }%
  \textbf{\bibinfo {volume} {22}},\ \bibinfo {pages} {363} (\bibinfo {year}
  {1980})\BibitemShut{NoStop}%
\bibitem{Furukawa:82}%
  \BibitemOpen
  \bibfield{author}{%
  \bibinfo {author} {\bibfnamefont{H.}~\bibnamefont{Furukawa}}\ and\ \bibinfo
  {author} {\bibfnamefont{K.}~\bibnamefont{Binder}},\ }%
  \bibfield{journal}{%
  \bibinfo {journal} {Phys.\ Rev.\ A}\ }%
  \textbf{\bibinfo {volume} {26}},\ \bibinfo {pages} {556} (\bibinfo {year}
  {1982})\BibitemShut{NoStop}%
\bibitem{Biskup:02}%
  \BibitemOpen
  \bibfield{author}{%
  \bibinfo {author} {\bibfnamefont{M.}~\bibnamefont{Biskup}}, \bibinfo {author}
  {\bibfnamefont{L.}~\bibnamefont{Chayes}},\ and\ \bibinfo {author}
  {\bibfnamefont{R.}~\bibnamefont{Koteck{\'{y}}}},\ }%
  \bibfield{journal}{%
  \bibinfo {journal} {EPL}\ }%
  \textbf{\bibinfo {volume} {60}},\ \bibinfo {pages} {21} (\bibinfo {year}
  {2002})\BibitemShut{NoStop}%
\bibitem{Biskup:03}%
  \BibitemOpen
  \bibfield{author}{%
  \bibinfo {author} {\bibfnamefont{M.}~\bibnamefont{Biskup}}, \bibinfo {author}
  {\bibfnamefont{L.}~\bibnamefont{Chayes}},\ and\ \bibinfo {author}
  {\bibfnamefont{R.}~\bibnamefont{Koteck{\'{y}}}},\ }%
  \bibfield{journal}{%
  \bibinfo {journal} {Commun.\ Math.\ Phys.}\ }%
  \textbf{\bibinfo {volume} {242}},\ \bibinfo {pages} {137} (\bibinfo {year}
  {2003})\BibitemShut{NoStop}%
\bibitem{Neuhaus:03}%
  \BibitemOpen
  \bibfield{author}{%
  \bibinfo {author} {\bibfnamefont{T.}~\bibnamefont{Neuhaus}}\ and\ \bibinfo
  {author} {\bibfnamefont{J.~S.}\ \bibnamefont{Hager}},\ }%
  \bibfield{journal}{%
  \bibinfo {journal} {J.\ Stat.\ Phys.}\ }%
  \textbf{\bibinfo {volume} {113}},\ \bibinfo {pages} {47} (\bibinfo {year}
  {2003})\BibitemShut{NoStop}%
\bibitem{Biskup:04}%
  \BibitemOpen
  \bibfield{author}{%
  \bibinfo {author} {\bibfnamefont{M.}~\bibnamefont{Biskup}}, \bibinfo {author}
  {\bibfnamefont{L.}~\bibnamefont{Chayes}},\ and\ \bibinfo {author}
  {\bibfnamefont{R.}~\bibnamefont{Koteck{\'{y}}}},\ }%
  \bibfield{journal}{%
  \bibinfo {journal} {J.\ Stat.\ Phys.}\ }%
  \textbf{\bibinfo {volume} {116}},\ \bibinfo {pages} {175} (\bibinfo {year}
  {2004})\BibitemShut{NoStop}%
\bibitem{MacDowell:04}%
  \BibitemOpen
  \bibfield{author}{%
  \bibinfo {author} {\bibfnamefont{L.~G.}\ \bibnamefont{MacDowell}}, \bibinfo
  {author} {\bibfnamefont{P.}~\bibnamefont{Virnau}}, \bibinfo {author}
  {\bibfnamefont{M.}~\bibnamefont{M{\"u}ller}},\ and\ \bibinfo {author}
  {\bibfnamefont{K.}~\bibnamefont{Binder}},\ }%
  \bibfield{journal}{%
  \bibinfo {journal} {J.\ Chem.\ Phys.}\ }%
  \textbf{\bibinfo {volume} {120}},\ \bibinfo {pages} {5293} (\bibinfo {year}
  {2004})\BibitemShut{NoStop}%
\bibitem{Nussbaumer:06}%
  \BibitemOpen
  \bibfield{author}{%
  \bibinfo {author} {\bibfnamefont{A.}~\bibnamefont{Nu{\ss}baumer}}, \bibinfo
  {author} {\bibfnamefont{E.}~\bibnamefont{Bittner}}, \bibinfo {author}
  {\bibfnamefont{T.}~\bibnamefont{Neuhaus}},\ and\ \bibinfo {author}
  {\bibfnamefont{W.}~\bibnamefont{Janke}},\ }%
  \bibfield{journal}{%
  \bibinfo {journal} {EPL}\ }%
  \textbf{\bibinfo {volume} {75}},\ \bibinfo {pages} {716} (\bibinfo {year}
  {2006})\BibitemShut{NoStop}%
\bibitem{Nussbaumer:10}%
  \BibitemOpen
  \bibfield{author}{%
  \bibinfo {author} {\bibfnamefont{A.}~\bibnamefont{Nu{\ss}baumer}}, \bibinfo
  {author} {\bibfnamefont{E.}~\bibnamefont{Bittner}},\ and\ \bibinfo {author}
  {\bibfnamefont{W.}~\bibnamefont{Janke}},\ }%
  \bibfield{journal}{%
  \bibinfo {journal} {Prog.\ Theor.\ Phys.\ Suppl.}\ }%
  \textbf{\bibinfo {volume} {184}},\ \bibinfo {pages} {400} (\bibinfo {year}
  {2010})\BibitemShut{NoStop}%
\bibitem{Bauer:10}%
  \BibitemOpen
  \bibfield{author}{%
  \bibinfo {author} {\bibfnamefont{B.}~\bibnamefont{Bauer}}, \bibinfo {author}
  {\bibfnamefont{E.}~\bibnamefont{Gull}}, \bibinfo {author}
  {\bibfnamefont{S.}~\bibnamefont{Trebst}}, \bibinfo {author}
  {\bibfnamefont{M.}~\bibnamefont{Troyer}},\ and\ \bibinfo {author}
  {\bibfnamefont{D.~A.}\ \bibnamefont{Huse}},\ }%
  \bibfield{journal}{%
  \bibinfo {journal} {J.\ Stat.\ Mech.},\ \bibinfo {pages} {P01020}}%
   (\bibinfo {year} {2010})\BibitemShut{NoStop}%
\bibitem{Nogawa:11}%
  \BibitemOpen
  \bibfield{author}{%
  \bibinfo {author} {\bibfnamefont{T.}~\bibnamefont{Nogawa}}, \bibinfo {author}
  {\bibfnamefont{N.}~\bibnamefont{Ito}},\ and\ \bibinfo {author}
  {\bibfnamefont{H.}~\bibnamefont{Watanabe}},\ }%
  \bibfield{journal}{%
  \bibinfo {journal} {Phys.\ Rev.\ E}\ }%
  \textbf{\bibinfo {volume} {84}},\ \bibinfo {pages} {061107} (\bibinfo {year}
  {2011})\BibitemShut{NoStop}%
\bibitem{Meisl:14}%
  \BibitemOpen
  \bibfield{author}{%
  \bibinfo {author} {\bibfnamefont{G.}~\bibnamefont{Meisl}}, \bibinfo {author}
  {\bibfnamefont{X.}~\bibnamefont{Yang}}, \bibinfo {author}
  {\bibfnamefont{E.}~\bibnamefont{Hellstrand}}, \bibinfo {author}
  {\bibfnamefont{B.}~\bibnamefont{Frohm}}, \bibinfo {author}
  {\bibfnamefont{J.~B.}\ \bibnamefont{Kirkegaard}}, \bibinfo {author}
  {\bibfnamefont{S.}~\bibnamefont{Cohen}}, \bibinfo {author}
  {\bibfnamefont{C.~M.}\ \bibnamefont{Dobson}}, \bibinfo {author}
  {\bibfnamefont{S.}~\bibnamefont{Linse}},\ and\ \bibinfo {author}
  {\bibfnamefont{T.~P.~J.}\ \bibnamefont{Knowles}},\ }%
  \bibfield{journal}{%
  \bibinfo {journal} {Proc.\ Natl.\ Acad.\ Sci.\ USA}\ }%
  \textbf{\bibinfo {volume} {111}},\ \bibinfo {pages} {9384} (\bibinfo {year}
  {2014})\BibitemShut{NoStop}%
\bibitem{Walsh:97}%
  \BibitemOpen
  \bibfield{author}{%
  \bibinfo {author} {\bibfnamefont{D.~M.}\ \bibnamefont{Walsh}}, \bibinfo
  {author} {\bibfnamefont{A.}~\bibnamefont{Lomakin}}, \bibinfo {author}
  {\bibfnamefont{G.~B.}\ \bibnamefont{Benedek}}, \bibinfo {author}
  {\bibfnamefont{M.~M.}\ \bibnamefont{Condron}},\ and\ \bibinfo {author}
  {\bibfnamefont{D.~B.}\ \bibnamefont{Teplow}},\ }%
  \bibfield{journal}{%
  \bibinfo {journal} {J.\ Biol.\ Chem.}\ }%
  \textbf{\bibinfo {volume} {272}},\ \bibinfo {pages} {22364} (\bibinfo {year}
  {1997})\BibitemShut{NoStop}%
\end{thebibliography}

%

\end{document}